\documentclass[a4paper,11pt]{article}
\pdfoutput=1 

\usepackage{jheppub} 

\usepackage{commath}
\usepackage{amsthm,amssymb,amsmath,mathrsfs}
\usepackage{feynmf}
\def\be {\begin{equation}}
\def\ee {\end{equation}}
\def\bea {\begin{eqnarray}}
\def\eea {\end{eqnarray}}
\def\bc {\begin{center}}
\def\ec {\end{center}}
\def\bfg {\begin{figure}}
\def\efg {\end{figure}}
\def\bi {\begin{itemize}}
\def\ei {\end{itemize}}

\def\le {\left}
\def\ri {\right}
\def\p {\partial}

\def\a  {\alpha}

\def\d  {\delta}

\def\e  {\eta}
\def\m  {\mu}
\def\n  {\nu}
\def\C  {\Gamma}

\newcommand{\sd}[2]{\overset{(s)}{#1}{}_{#2}}

\title{\boldmath
Continuous-spin field propagator and interaction with matter
}

\author
{Xavier Bekaert$^{a,b}$,}
\affiliation
{$^a$ Laboratoire de Math\'ematiques et Physique Th\'eorique\\
Unit\'e Mixte de Recherche $7350$ du CNRS\\
F\'ed\'eration de Recherche $2964$ Denis Poisson\\
Universit\'e Fran\c{c}ois Rabelais, Parc de Grandmont\\
37200 Tours, France}
\affiliation
{$^b$ B.W. Lee Center for Fields, Gravity and Strings\\ Institute for Basic Science\\ Daejeon, Korea}
\author
{Jihad Mourad$^{c}$}
\affiliation
{$^c$ AstroParticule et Cosmologie\\
Unit\'e Mixte de Recherche $7164$ du CNRS,\\
Universit\'e Paris VII, B\^atiment Condorcet\\
75205 Paris Cedex 13, France}
\author
{and Mojtaba Najafizadeh$^{a,d}$}
\affiliation
{$^d$ School of Physics\\  Institute for Research in Fundamental Sciences (IPM) \\ P.O.Box 19395-5531, Tehran, Iran }
\emailAdd{xavier.bekaert@lmpt.univ-tours.fr}
\emailAdd{mourad@apc.univ-paris7.fr}
\emailAdd{mojtaba.najafizadeh@lmpt.univ-tours.fr}

\abstract{Action principles for the single and double valued continuous-spin representations of the Poincar\'e group have been recently proposed in a Segal-like formulation. We address three related issues:
First, we explain how to obtain these actions directly from the Fronsdal-like and Fang-Fronsdal-like equations by solving the traceless constraints in Fourier space.
Second, we introduce a current, similar to the one of Berends, Burgers and Van Dam, which is bilinear in a pair of scalar matter fields, to which the bosonic continuous-spin field can couple minimally.
Third, 
we investigate 
the current exchange mediated by a continuous-spin particle obtained from this action principle and investigate whether it propagates the right degrees of freedom, and whether it reproduces the known result for massless higher-spin fields in the helicity limit.
}


\begin{document}

\maketitle

\pagebreak


\section{Introduction}

Wigner showed that the unitary irreducible representations (UIRs) of the Poincar\'e group are determined by those of the corresponding little group \cite{Wigner}. In $D$ spacetime dimensions, massive particles are determined by representations of the rotation group $SO(D-1)$ and massless ones by representations of the Euclidean group $ISO(D-2)$ of the transverse plane ${\mathbb R}^{D-2}$.

Massive particles in $D=4$ spacetime dimensions\footnote{The lectures \cite{Bekaert:2006py} provide a self-contained introduction to the classification of the unitary irreducible representations of the Poincar\'e group in any dimension. In the present paper, we will stick to $D=4$ for the sake of definiteness but, for totally symmetric tensor(-spinor) representations, our results have straightforward generalisations to any dimension $D>2$.}
are labeled by their spin $s\in\frac12\mathbb N$ defined, more covariantly, via the eigenvalue of the quartic Casimir operator
\be
\hat{W}^2=-\,m^2 \,s(s+D-3)  \label{W^2}
\ee
where $\hat{W}^\m= \frac{1}{2} \, \epsilon^{\m\n \rho\sigma} \, \hat{p}_\n \, \hat{J}_{\rho\sigma}$ is the Pauli-Lubanski vector. The mass squared is defined as the eigenvalue of the quadratic Casimir operator, $\hat{p}^2=m^2$, in the mostly minus signature.

Massless particles are divided into two different types of representations. The well-known ones have two polarisations in four spacetime dimensions and are labeled by the eigenvalues of the helicity operator $\hat{h}=\vec{S} \cdot \frac{\vec{p}}{|\vec{p}|}$ where $\vec{S}$ is the spin operator.
These representations have vanishing eigenvalue of the quartic Casimir operator in any dimension and are usually called ``helicity'' representations, although they could also be referred to as ``discrete-spin'' (or ``finite-spin'') representations.
Another, less studied, kind of massless particles are determined by a continuous energy scale $\m$ via the eigenvalue of the quartic Casimir operator
\be
\hat{W}^2=-\, \label{mu^2}
\m^2
\ee
and their spectrum comprises infinitely many polarisations which mix under Lorentz boosts. These representations are often called ``continuous-spin'' representations \cite{Wigner} or ``infinite-spin'' ones \cite{Wigner 2}.\footnote{See \cite{Bekaert:2017khg} for a recent review on these representations.}
When $\mu=0$, the helicity eigenstates do not mix while they do when $\mu\neq 0$, so the continuous parameter $\m$ controls the degree of mixing.
In fact, in the ``helicity limit'' $\m\to 0$, the continuous-spin representation becomes reducible and decomposes into the direct sum of all (integer vs half-integer) helicity representations (respectively for the single vs double valued representation).
As noticed by Khan and Ramond \cite{KR},
the continuous-spin representation can be obtained as the limit
\be
m\rightarrow 0\,, \quad  s \rightarrow \infty\,,  \quad ms\,=\,\m\,=\,\mbox{constant}\,, \label{Limit}
\ee
of massive higher-spin representations.
In particular, the limit of (\ref{W^2}) indeed gives (\ref{mu^2}). This limit provides a suggestive physical picture of a continuous-spin particle (CSP) as a massive higher-spin particle (HSP) with very large spin $s\gg 1$ but mass much below the relevant energy scale $E$ of the process: $m=\mu/s\ll E$.

For the helicity representations, Fronsdal and Fang-Fronsdal equations are gauge-invariant equations of motion for, respectively, integer and half-integer spin fields with trace constraints on the gauge fields and parameters \cite{F 1, F 2}. For the continuous-spin representations, Fronsdal-like and Fang-Fronsdal-like equations  were obtained in \cite{BM} from massive higher-spin field equations by taking the limit \eqref{Limit} (and extracting a suitable divergent factor of the wavefunction).
These equations provide a gauge formulation of the bosonic and fermionic continuous-spin field with constraints.
Recently, a local gauge-invariant action for \textit{bosonic} HSP\footnote{In their higher-spin formalism, the auxiliary vector is restricted to live on a hyperboloid, like what was considered for CSPs, while for HSPs a restriction to the null cone is actually more natural in order to obtain the double trace constraint on the gauge field and the tracelessness of the gauge parameter in dual space, as will be explained here (see also \cite{Rivelles:2014fsa,Rivelles:2016rwo} for a similar point of view).} and CSP with unconstrained gauge field and parameter was proposed by Schuster and Toro \cite{ST PRD}
and analysed further in \cite{Rivelles:2014fsa,Rivelles:2016rwo}. This unconstrained formulation is closely related to the action principle proposed by Segal \cite{Segal:2001qq} for bosonic higher-spin fields on (anti) de Sitter spacetime, since they coincide in the respective helicity or flat limit.
A local gauge-invariant action for the \textit{fermionic} CSPs without any constraint on the gauge field and parameter was proposed in \cite{BNS}. Recently, local actions \`a la Fronsdal for bosonic and fermionic particles with infinitely many spinning degrees of freedom (such as the continuous-spin and the tachyonic representations of the Poincar\'e group with non-zero spin) propagating on Minkowski and (anti) de Sitter backgrounds were proposed by Metsaev in \cite{Metsaev:2016lhs,Metsaev:2017ytk}.\footnote{See also \cite{Zinoviev:2017rnj} for a frame-like approach to the action principles of CSP.} Contact was made between the two approaches in \cite{Najafizadeh:2017tin}.

The first goal of this paper is to explain how to derive the actions of bosonic and fermionic CSP gauge fields proposed in \cite{ST PRD, BNS} directly from the
Fronsdal-like and Fang-Fronsdal-like equations \cite{BM} by solving the trace-like constraints in the auxiliary Fourier space.\footnote{The fermionic action \cite{BNS} was obtained with this procedure, but the technical details were not spelled out in this letter.} This includes HSP results from CSP ones by taking the appropriate helicity limit. The second goal of this work is to discuss the couplings to scalar matter currents. The third goal is to
investigate whether the correct propagator of a bosonic CSP is obtained from the bosonic action \cite{ST PRD}.

The layout of this paper is as follows. In section \ref{s2}, we will derive the Schuster-Toro action \cite{ST PRD} from the Fronsdal-like equation \cite{BM} by solving the (double-)trace constraints and by noticing that the resulting equation possesses a hermitian kinetic operator. In section \ref{s3}, we will reach the fermionic CSP action \cite{BNS} from the Fang-Fronsdal-like equation \cite{BM} by following the same method as in the previous section. In section \ref{s4}, we review the relation between the (Fang-)Fronsdal-like equation \cite{BM} and the Wigner equations \cite{wi} for the single (double) valued continuous-spin representation. In section \ref{s5}, we derive conservation-like conditions for currents to couple consistently with the constrained gauge fields, we show how to obtain them from the infinite-spin limit \eqref{Limit} and show that they are equivalent to the conditions in \cite{ST PRD,BNS}. In section \ref{s6}, we provide a current which obeys this conservation condition and which is built from two distinct scalar fields, thereby providing the first example of local cubic coupling of a CSP to matter. In section \ref{s7}, we obtain the residue of the propagator for a bosonic CSP from the Schuster-Toro action and discuss whether it gives the correct physical result (in the sense that the unphysical trace-like part of the current should not propagate). The conclusions and perspectives for future work are displayed in section \ref{s8}. The appendices gather diverse material. Our conventions have been placed inside the appendix \ref{appA}, while the appendix \ref{Fermionic CSP equation} provides a review on the fermionic CSP equations. Useful results on the integrals over the auxiliary variable can be found in the appendix \ref{eta integrals}. The formulas of current exchange in the Euclidean signature are derived in the appendix \ref{eta_integrals}. These two appendices make use of standard facts about special functions summarised in the appendix \ref{useful}.

\section{Bosonic higher-spin and continuous-spin gauge fields}\label{s2}

Introducing an auxiliary vector $\omega^\mu$, we can pack a collection of totally symmetric tensor fields $\Phi_{\m_{1}\ldots\m_{s}}(x)$ of all integer rank $s\in\mathbb N$ into a single generating function
\be
\Phi(x,\omega)=\sum\limits_{s=0}^\infty\frac{1}{s!} \, \, \omega^{\m_{1}}   \ldots \omega^{\m_{s}} \, \,
\Phi_{\m_{1}\ldots\m_{s}}(x)\,.  \label{generating func.}
\ee
The higher-spin and continuous-spin equations \`a la Fronsdal are given by \cite{BM}
\be
 \left[ -\,\Box_x + \left(\omega \cdot \p_x + i\, \sigma \m\right) \left(\p_{\omega} \cdot \p_x  \right)
 -
 {\frac{1}{2}}\,(\omega \cdot  \p_x + i \,\sigma \m)^2
\left( \p_{\omega} \cdot \p_\omega - \sigma \right)\right]\Phi(x,\omega)=0\,, \label{Bosonic}
\ee
where $\p_x=\frac{\p}{\p x}$ , $\partial_{\omega}=\frac{\partial}{\partial \omega}$ and
\be
\sigma = \left\{
           \begin{array}{ll}
             0 & \hbox{for higher-spin,} \\
             1 & \hbox{for continuous-spin.}
           \end{array}
         \right.  \label{sigma}
\ee
This equation is gauge invariant under the gauge transformation
\be
\delta_\epsilon \Phi (x,\omega)\,=\,\left(\omega \cdot \p_x + i\, \sigma \m \right)\, \epsilon (x,\omega)\,,   \label{Bosonic GT}
\ee
with the trace condition on $\epsilon$
\be
\left( \p_\omega \cdot \p_\omega - \sigma \right) \, \epsilon (x,\omega) = 0\,,   \label{Bosonic GP constraint}
\ee
where $\epsilon$ is a gauge transformation parameter,
\be
\epsilon(x,\omega)=\sum\limits_{s=1}^\infty\tfrac{1}{(s-1)!} \, \, \omega^{\m_{1}}   \ldots \omega^{\m_{s-1}} \, \,
\epsilon_{\m_{1}\ldots\m_{s-1}}(x)\,.
\ee
The double-trace constraint on the gauge field reads
\be
\left( \p_\omega \cdot \p_\omega - \sigma \right)^2 \, \Phi (x,\omega) = 0\,.   \label{Bosonic F constraint}
\ee

\subsection{Equations in Fourier space}

To obtain the Euler-Lagrange equation in \cite{ST PRD}, we should work in $\e$-space, the dual space of the $\omega$-space:
\be
\widetilde{\Phi} (x,\e) = \int \frac{d^D\omega}{(2\pi)^\frac{D}2}\,\exp(\,-\,i\,\eta\cdot\omega\,)\,\Phi (x,\omega)\,.
\ee
Writing (\ref{Bosonic F constraint}) in $\e$-space,
we find
\be
 \left( \e^2 + \sigma \right)^2 \, \widetilde{\Phi} (x,\e) = 0\,, 
\ee
whose general solution is
\be
\widetilde{\Phi} (x,\e) =  \delta'(\eta^2 +\sigma) \, \Phi (x,\e)\,, \label{Bosonic F constraint eta}
\ee
where $\Phi(x,\e)$ is an arbitrary function which is unconstrained, unlike $\Phi(x,\omega)$. The equation \eqref{Bosonic F constraint eta} shows that dynamical fields live on the hypersurface $\,\eta^2+\sigma=0$ in the $\eta$-space.
Notice that there are an infinite number of fields ${\Phi} (x,\e)$ corresponding to the same given $\widetilde\Phi(x,\e)$. More precisely, the arbitrariness is encoded into the gauge symmetry
\be
\delta_\chi \Phi(x,\e) = \le(\e^2 + \sigma \ri)^2 \, \chi(x,\e)\,, \label{kapa transformation}
\ee
where $\chi$ is an arbitrary unconstrained function.
The equation (\ref{Bosonic}) in dual space becomes
\be
 \left[ -\,\Box_x - \left(\p_{\e} \cdot \p_x + \sigma \m\right) \left(\e \cdot \p_x  \right)
 -
 {1\over 2}\, \left(\p_{\e} \cdot  \p_x + \sigma \m \right)^2
\left( \e^2 + \sigma \right)\right]\widetilde{\Phi}(x,\e)=0\,, \label{Bosonic eta}
\ee
which, using (\ref{Bosonic F constraint eta}), reduces to
\be
\left[ \,-\,  \d'(\e^2 +\sigma) \, \Box_x
+ \frac{1}{2}\left( \p_{\e} \cdot \p_x + \sigma \m \right) \d(\e^2 +\sigma) \left( \p_{\e} \cdot \p_x + \sigma \m \right) \,\right] \Phi(x,\e) =0\,. \label{Bosonic eta equation}
\ee
For $\sigma=1$, this equation of motion is the one from \cite{ST PRD} but we obtained it from the Fronsdal-like equation (\ref{Bosonic}) in \cite{BM} by solving the double-trace condition. This equation of motion is invariant under the gauge transformation \eqref{kapa transformation}.
To obtain the remaining gauge transformation we should write (\ref{Bosonic GP constraint}) in $\e$-space, $\left( \e^2 + \sigma \right) \widetilde{\epsilon} (x,\e)=0$, then we find the complete set of the gauge transformations in \cite{ST PRD}
\be
\widetilde{\epsilon} (x,\e) =  \delta(\eta^2 +\sigma)\, \epsilon (x,\e)\,, \label{Bosonic GP constraint eta}
\ee
where $\epsilon (x,\e)$ is an arbitrary unconstrained function.
Considering (\ref{Bosonic GT}) in dual space
\be
\delta_\epsilon \widetilde{\Phi} (x,\e)\,=\,\left(\p_{\e} \cdot \p_x + \sigma \m \right)\, \widetilde{\epsilon} (x,\e)\,,   \label{Bosonic GT eta}
\ee
and putting (\ref{Bosonic GP constraint eta}) and (\ref{Bosonic F constraint eta}) into it, we  find
\be
 \delta_{\epsilon,\chi} \Phi (x,\e)\,= \left[    \,\e \cdot \p_x  -  \frac{1}{2} \left(\e^2+\sigma  \right)   \left(\p_\e \cdot \p_x +  \sigma \m      \right) \right] \epsilon (x,\e)
 + \frac{1}{2} \le(\e^2 + \sigma \ri)^2 \, \chi(x,\e)\,,
 \label{Bosonic GT eta 1}
\ee
where the gauge parameters have been rescaled by a constant factor to match the normalisation in \cite{ST PRD}.

\subsection{Bosonic action}

Considering the equation of motion (\ref{Bosonic eta equation}) for the redefined gauge field, we can introduce a kinetic operator as
\be
\widehat{K}(\partial_x,\e,\partial_\eta)=- \, \d'(\e^2 +\sigma) \, \Box_x + \frac{1}{2} \left( \p_{\e} \cdot \p_x + \sigma \m \right)  \d(\e^2 +\sigma) \left( \p_{\e} \cdot \p_x + \sigma \m \right)
\ee
 such that $\widehat{K} \Phi= 0$.
One can check that the kinetic operator is Hermitian,
\be \widehat{K}^ \dag = \widehat{K}\,, \ee
with respect to the Hermitian conjugation
\be
 (\p_x )^\dag \equiv - \, \p_x \,, \quad \quad (\p_{\e} )^\dag \equiv - \, \p_{\e} \,,\quad \quad {\e}^\dag \equiv {\e} \,, \label{rond}
 \ee
This property is highly remarkable because the original Fronsdal-like kinetic operator in \eqref{Bosonic} is not Hermitian. In fact, in the Fronsdal formulation the corresponding operator needs to be completed by a term involving the trace, thereby generalising the completion of the Ricci tensor to the Einstein tensor in the spin-two case.
This suggests writing the free action as
\bea
 S_{free}  & = & \frac{1}{2} \, { \int  d^4 x \, d^4 \eta }  \,\, \Phi(x,\e) ~ \widehat{K}(\partial_x,\e,\partial_\eta)~ \Phi(x,\e)  \label{Bosonic Action} \\
                & = & \frac{1}{2} \, { \int  d^4 x \, d^4 \eta }  \,\, \Phi \le[ {- \,\d'(\e^2 +\sigma) \, \Box_x + \tfrac{1}{2} \left( \p_{\e} \cdot \p_x + \sigma \m \right)  \d(\e^2 +\sigma) \left( \p_{\e} \cdot \p_x + \sigma \m \right)}\ri] \Phi\nonumber \,,
\eea
 which is real and invariant under the gauge transformation (\ref{Bosonic GT eta 1}).
This action introduces a gauge field theory of higher-spin $(\sigma=0)$ and continuous-spin $(\sigma=1)$ particles which, upon integration by part, reproduces the Schuster-Toro action \cite{ST PRD}.
It has been obtained straightforwardly from the Fronsdal-like equation and this method will be applied to derive the fermionic action \cite{BNS} in the next section.

\section{Fermionic higher-spin and continuous-spin gauge fields}\label{s3}

We consider a collection of totally symmetric spinor-tensor fields $\Psi_{\m_{1}\ldots\m_{n}}(x)$ of all half-integer spin $s=n+\frac{1}{2}$, which we pack in the generating function
\be
\Psi(x,\omega)=\sum\limits_{n=0}^\infty\,\frac{1}{n!} \, \, \omega^{\m_{1}}   \ldots \omega^{\m_{n}} \, \,
\Psi_{\m_{1}\ldots\m_{n}}(x),  \label{psi}
\ee
where the spinor index is left implicit.
The Fang-Fronsdal equation \cite{F 2} and Fang-Fronsdal-like equation \cite{BM}, reviewed in (\ref{HS I 1 1}), are given by (respectively for $\sigma=0$ and $\sigma=1$)
\be
\Big[\gamma\cdot \p_x - \le(\omega\cdot \p_x + i\, \sigma \m \ri)  \le( \gamma\cdot{\partial_{\omega} + \sigma } \ri)\Big]\Psi(x,\omega) =0\,, \label{Fermionic}
\ee
where $\gamma^\m$ are gamma matrices in 4 dimensions.
This equation is gauge invariant under the transformations
\be
\delta_\varepsilon \Psi(x,\omega)=\big(\omega \cdot \p_x + i\,\sigma\mu\big)\,\varepsilon(x,\omega)\,,  \label{Fermionic GT}
\ee
where $\varepsilon$ is a spinor-tensor gauge  parameter
\be
\varepsilon(x,\omega)=\sum\limits^\infty_{n=1}\tfrac{1}{(n-1)!} \, \, \omega^{\m_{1}}   \ldots \omega^{\m_{n-1}}\, \,
\varepsilon_{\m_{1}\ldots\,\m_{n-1}}(x),  \label{epsilon}
\ee
which is constrained by the gamma-trace conditions
\be
\left(\gamma\cdot \partial_{\omega} + \sigma\right)\,\varepsilon(x,\omega)=0\,. \label{Fermionic GP constraint}
\ee
The triple gamma-trace condition can be expressed as
\be
\left(\gamma\cdot\partial_{\omega} + \sigma\right)\left(
\partial_{\omega}\cdot \partial_{\omega} - \sigma\right)\,\Psi(x,\omega)=0\,. \label{Fermionic F constraint}
\ee

\subsection{Equations in Fourier space}

To cancel the constraints on the gauge field and parameter, we go to the Fourier space of $\omega$.
%
In this space, the equation (\ref{Fermionic F constraint}) becomes
\be
\left(\gamma\cdot\eta - i \, \sigma \right)\left(
\eta^2 + \sigma\right)\widetilde{\Psi}(x, \e)=0\,.   \label{Fermionic F constraint eta}
\ee
Multiplying this relation by $( \gamma\cdot {\eta} + i \,\sigma)$, we find\footnote{Due to \eqref{sigma}, in the sections \ref{s3} and \ref{s5}, for simplicity, we applied $\sigma^2=\sigma$\,.}
\be
\big( \e^2 + \sigma \big)^2 \, \widetilde\Psi(x,\e) =0\,.
\ee
The general solution reads
\be
 \widetilde\Psi(x,\e) = \d'(\e^2 +\sigma) \,\zeta(x,\e)\,, \label{chi}
\ee
where $\zeta$ is an arbitrary function. Putting (\ref{chi}) into (\ref{Fermionic F constraint eta}) we can deduce
\be
\zeta(x,\e)= \big( \gamma\cdot {\eta} + i \,\sigma \big) \Psi(x,\e)
\ee
where $\Psi(x,\e)$ is a new arbitrary function. Finally, we find
\be
\widetilde\Psi(x,\eta)= \delta'(\eta^2 +\sigma) \big( \gamma\cdot {\eta}+ i\,\sigma  \big) \Psi(x,\eta)\,.
 \label{Fermionic psi tilde}
\ee

To find the gauge transformation parameter in $\e$-space, we write (\ref{Fermionic GP constraint}) as
\be
\left(\gamma\cdot \e - i\,\sigma\right)\widetilde\varepsilon(x,\e)=0\,, \label{Fermionic varepsilon tilde}
\ee
and multiplying by $(\gamma\cdot \e + i\,\sigma)$, we obtain
\be
\big( \e^2 + \sigma \big) \, \widetilde\varepsilon(x,\e)=0\,.
\ee
This means that we can consider $\widetilde\varepsilon$ as
\be
\widetilde\varepsilon(x,\e) = \d(\e^2 + \sigma) \,\xi(x,\e)\,, \label{Fermionic epsilon tilde 1}
\ee
where $\xi$ is an arbitrary function. Using (\ref{Fermionic varepsilon tilde}) and (\ref{Fermionic epsilon tilde 1}) we find
\be
 \xi(x,\e)=  (\gamma\cdot \e + i\,\sigma) \,\boldsymbol{\epsilon}(x,\e)
\ee
and then we reach
\be
\widetilde\varepsilon(x,\e) = \d(\e^2 + \sigma) (\gamma\cdot \e + i\,\sigma) \,\boldsymbol{\epsilon}(x,\e) \,, \label{Fermionic epsilon tilde}
\ee
where $\boldsymbol{\epsilon}(x,\e)$ is an unconstrained gauge parameter.

Writing (\ref{Fermionic GT}) in $\e$-space
\be
\delta_{\widetilde\varepsilon} \widetilde\Psi(x,\e)=\big(\p_\e \cdot \p_x + \sigma \mu\big)\,\widetilde\varepsilon(x,\e)\,,  \label{Fermionic GT eta 1}
\ee
and substituting (\ref{Fermionic psi tilde}) and (\ref{Fermionic epsilon tilde}) into it, we find
\be
\delta_{\boldsymbol{\epsilon}} \Psi(x,\e)=\Big[\le( \gamma \cdot \p_x \ri) \le(\gamma\cdot \eta + i \,\sigma  \ri) - \le(\eta^2 +\sigma \ri) \le({\partial_{\eta}} \cdot \p_x + \sigma \m \ri) \Big]  \,\boldsymbol{\epsilon}(x,\e) \,.  \label{epsilon symm}
\ee
%
The equation of motion (\ref{Fermionic}) in Fourier space reads
\be
\Big[\gamma\cdot \p_x + \le(\p_\e \cdot \p_x +  \sigma \m \ri)  \le( \gamma\cdot \e - i\,\sigma \ri)\Big] \widetilde\Psi(x,\e) =0\,, \label{fermionic EOM 1}
\ee
which, using (\ref{Fermionic psi tilde}), becomes
\be
\Big[\delta'(\eta^2 + \sigma) \big( \gamma\cdot {\eta} -  i\,\sigma  \big) (\gamma  \cdot \p_x)
+ \delta(\eta^2 + \sigma) \big({\partial_{\eta}} \cdot \p_x + \sigma \m \big)  \Big] \Psi (x,\eta)=0  \,. \label{Fermionic EOM}
\ee
This is the equation of motion for fermionic higher-spin ($\sigma=0$) and continuous-spin ($\sigma=1$) fields in an unconstrained formulation where the trace-like constraints have been solved algebraically in $\eta$-space. In the CSP case, one reads exactly the equation of motion of \cite{BNS}. This equation is invariant under the extra gauge transformation
\be
\delta_{\boldsymbol{\chi}} \Psi(x,\e)= (\e^2+\sigma) \le( \gamma\cdot {\eta} -  i\,\sigma  \ri) \, \boldsymbol{\chi}(x,\e)\,,  \label{chi symm}
\ee
related to the arbitrariness in \eqref{Fermionic psi tilde}, where $\boldsymbol{\chi}$ is an unconstrained arbitrary spinor gauge parameter. Therefore, the equation of motion (\ref{Fermionic EOM}) is invariant under two gauge transformations
\be
\delta_{\boldsymbol{\epsilon}, \boldsymbol{\chi}} \Psi(x,\e) = \delta_{\boldsymbol{\epsilon}} \Psi(x,\e) + \delta_{\boldsymbol{\chi}} \Psi(x,\e)\,, \label{Fermionic GT eta}\,.
\ee
It is straightforward to show that multiplying the fermionic equation of motion (\ref{fermionic EOM 1}) by the operator
\be
\Big[\gamma\cdot \p_x + \le(\p_\e \cdot \p_x +  \sigma \m \ri)  \le( \gamma\cdot \e + i\,\sigma \ri)\Big] \label{squar of fermionic EOM}
\ee
gives the bosonic equation of motion (\ref{Bosonic eta}) for the tensor-spinor field $\Psi$.

\subsection{Fermionic action}

We can introduce the kinetic operator
\be
\widehat{\mathcal{K}} (x,\eta) = \delta'(\eta^2 + \sigma) \big( \gamma\cdot {\eta} -  i\,\sigma  \big) (\gamma  \cdot \p_x)
+ \delta(\eta^2 + \sigma) \big({\partial_{\eta}} \cdot \p_x + \sigma \m \big) \,,  \label{Fermionic kinetic}
\ee
such that \eqref{Fermionic EOM} reads $\widehat{\mathcal{K}}\, \Psi=0\,.$ Using (\ref{rond}), it can be checked that
\be
\widehat{\mathcal{K}}^\dag = \gamma^0 \, \widehat{\mathcal{K}} \,  \gamma^0 \,,
\ee
which means we can obtain our equation of motion (\ref{Fermionic EOM}) from the action
\bea
 S_{free}  & = &   { \int  d^4 x \, d^4 \eta }  \,\, \overline{\Psi} ~ \widehat{\mathcal{K}}~ \Psi \label{Fermionic Action} \\
                & = &   {  \int  d^4 x \, d^4 \eta }  \,\, \overline{\Psi}\,
 \Big[   \delta'(\eta^2 + \sigma) \big( \gamma\cdot {\eta} -  i\,\sigma  \big) (\gamma  \cdot \p_x)
+ \delta(\eta^2 + \sigma) \big({\partial_{\eta}} \cdot \p_x + \sigma \m \big)  \Big]\,  \Psi \,,\nonumber
\eea
where $\overline{\Psi}= {\Psi}^\dag\, \gamma^0$. The action is gauge invariant under the gauge transformation (\ref{Fermionic GT eta}).
The proposed action in \cite{BNS} is equal to this action by choosing $\sigma=1$.
The actions for the integer (\ref{Bosonic Action}) and half integer (\ref{Fermionic Action}) spins comprise both higher-spin and continuous-spin cases depending on the value of $\sigma$ following the prescription \eqref{sigma}.

\section{Wigner equations}\label{s4}

The wave equations of Wigner for the bosonic and fermionic CSPs, carrying a unitary irreducible representation of the Poincar\'e group, respectively single and double valued, were presented in \cite{wi,Bargmann:1948ck}. These equations can be obtained as a consequence of the Fronsdal-like and Fang-Fronsdal-like equations after gauge fixing \cite{BM}.

For the bosonic CSP, consider (\ref{Bosonic eta}) and (\ref{Bosonic GT eta}) with $\sigma=1$. After gauge fixing and defining the gauge invariant field\footnote{See the procedure in \cite{BM} for more details.}
\be\label{hatPhi}
\widehat{\Phi} (p,\e) = \int \frac{d^D\eta}{(2\pi)^\frac{D}2}\,\frac{d^Dx}{(2\pi)^\frac{D}2}\, \exp[\,-\,i\,(p\cdot x+\eta\cdot\omega)\,]\,\,(p\cdot \omega-\mu)\,\Phi (x,\omega)\,,
\ee
we find exactly the single-valued Wigner equations \cite{Bargmann:1948ck}
\be
p^2\, \widehat\Phi=0 \,,  \quad  \quad \le( p \cdot \e\ri)\, \widehat\Phi=0 \,, \nonumber
\ee
\be
\le(\e^2+1\ri)\, \widehat\Phi=0 \,,  \quad\quad   \le( p \cdot \p_\e + i \,\m\ri)\, \widehat\Phi=0 \,,  \label{Wigner boson}
\ee

For the fermionic CSP, after fixing the gauge transformation (\ref{Fermionic GT eta 1}) and using the equation (\ref{fermionic EOM 1}) we arrive at
\be
\le(\gamma \cdot p\ri) \, \widehat\Psi=0\,,   \quad  \quad \le( p \cdot \e\ri)\, \widehat\Psi=0 \,,  \nonumber
\ee
\be
\le(\gamma \cdot \e - i\,\ri)\, \widehat\Psi=0\,,   \quad\quad   \le( p \cdot \p_\e + i\, \m\ri)\, \widehat\Psi=0 \,, \label{Wigner fermion}
\ee
where $\widehat\Psi$ is defined similarly to \eqref{hatPhi}. These are the double-valued Wigner equations, except the third one which should be multiplied by $\le(\gamma \cdot \e + i\,\ri)$ to get exactly the results of \cite{Bargmann:1948ck} (see the explanations in \cite{BM}).

Using these equations, it can be shown that the square of the Pauli-Lubanski vector $\hat{W}^\m$ acting on the bosonic and fermionic gauge fields is proportional to $\m^2$ which implies that the Pauli-Lubanski vector mixes the helicity eigenstates. To illustrate this, we consider the Pauli-Lubanski vector as $\hat{W}^\m=  \, \epsilon^{\m\n \rho\sigma} \, \hat{p}_\n \, \hat{J}_{\rho\sigma}$ where the Lorentz generators $\hat{J}^{\m\n}$  are
\be
\hat{J}^{\m\n}=\left\{
                 \begin{array}{ll}
                   x^{[\m}~ \hat{p}^{\n]}+i\, \e^{[\m}~ \p_\e^{\n]}, & \hbox{for bosons;} \\
                   x^{[\m}~ \hat{p}^{\n]}+i \, \e^{[\m}~ \p_\e^{\n]} +\frac{i}2  \gamma^{[\m} \gamma^{\n]}, & \hbox{for fermions.}
                 \end{array}
               \right.     \footnote{The brackets stand $a^{[\m}~ b^{\n]}=a^\m b^\n - a^\n b^\m $}
\ee
Then the square of the Pauli-Lubanski vector $\hat{W}^2$ (assuming $\hat{p}^2 = 0$) for both bosons and fermions reads
\be
\hat{W}^2=2\,(\hat{p} \cdot \e) (\hat{p} \cdot \p_\e) (\e \cdot \p_\e) - (\hat{p} \cdot \e)^2 \, (\p_\e \cdot \p_\e) - \e^2 (\hat{p} \cdot \p_\e )^2 \,.
\ee
Acting the above on the gauge fields and using the Wigner equations ( (\ref{Wigner boson}) and (\ref{Wigner fermion}) ) we find
\be
\le( \hat{W}^2 +\, \sigma \, \m^2 \, \ri) \widehat\Phi= 0\,, \quad \quad  \le( \hat{W}^2 +\, \sigma \, \m^2 \, \ri) \widehat\Psi= 0 \,.
\ee
This is in agreement with the fact that the field $\widehat\Phi$ (respectively, $\widehat\Psi$) carries an irreducible representation of the Poincar\'e group for a bosonic (respectively, fermionic) CSP.

\section{Conservation condition}\label{s5}

In this section, we present and obtain the conservation-like condition for continuous-spin particles via an infinite-spin limit \eqref{Limit}.

The higher-spin conservation condition (for the integer and half-integer spins) have been given in \cite{F 1} and \cite{F 2} in terms of the higher-spin generalisations of the (linearised) Einstein tensor. Introducing an auxiliary vector $u$, we can define
\be
j(x,u)=\frac{1}{s!} \, \, u^{\m_{1}}   \ldots u^{\m_{s}} \, \,
j_{\m_{1}\ldots\m_{s}}(x)\,,
\ee
\be
\Sigma(x,u)=\frac{1}{n!} \, \, u^{\m_{1}}   \ldots u^{\m_{n}} \, \,
\Sigma_{\m_{1}\ldots\m_{n}}(x)\,,
\ee
where $j_{\m_{1}\ldots\m_{s}}(x)$ is a bosonic background current of spin $s$ and $\Sigma_{\m_{1}\ldots\m_{s}}(x)$ is a fermionic one of spin $s=n+\frac{1}{2}$, sourcing the respective massless higher-spin equations
\be
 \left[ -\,\Box_x - \left(\p_{u} \cdot \p_x\right) \left(u \cdot \p_x  \right)
 -
 {1\over 2}\, \left(\p_{u} \cdot  \p_x  \right)^2
\, u^2 \right]\widetilde{\Phi}(x,u)=j(x,u)\,, \label{Bosonic CC 1}
\ee
\be
\Big[\,\Gamma\cdot \p_x + \le(\p_u \cdot \p_x \ri)  \le( \Gamma\cdot u\ri)\Big] \widetilde\Psi(x,u) =\Sigma(x,u)\,, \label{fermionic CC 1}
\ee
where the matrices $\Gamma$ satisfy the Clifford algebra.
These equations respectively imply the following conservation conditions (sometimes called ``Bianchi identities'' when they are concerned with the left-hand-side of the equations of motion):
\be
\le[ \p_u \cdot \p_x - \tfrac{1}{2} (u \cdot \p_x ) (\p_u \cdot \p_u ) \ri] j(x,u)=0\,, \label{Bosonic CC 1'}
\ee
\be
\le[ \le(\Gamma \cdot \p_u\ri) \le(\Gamma \cdot \p_x\ri) -  (u \cdot \p_x ) (\p_u \cdot \p_u ) \ri] \Sigma(x,u)=0\,. \label{Fermionic CC 1'}
\ee
The double-trace constraint on $j$ and triple gamma-trace constraint on $\Sigma$ take the following forms
\be
(\p_u \cdot \p_u)^2~ j(x,u)=0 \,,  \quad \quad  (\Gamma \cdot \p_u)(\p_u \cdot \p_u) \,\Sigma(x,u)=0\,. \label{B F constraints}
\ee
To obtain the massive equations of (\ref{Bosonic CC 1}) and (\ref{fermionic CC 1}), we consider the 5-dimensional analogue of the previous equations and consider a Kaluza-Klein mode with momentum equal to $m$ along the fifth direction,
 \be
\le[ \p_u \cdot \p_x - i\, m\, \p_v - \tfrac{1}{2} \le(u \cdot \p_x  + i\, m \,v\ri) \le(\p_u \cdot \p_u - \p_v ^2 \ri) \ri] j(x,u,v)=0\,, \label{Bosonic CC 2}
\ee
\be
\le[ \le(\Gamma \cdot \p_u - i\, \Gamma^5  \p_v \ri) \le( \Gamma \cdot \p_x +  m \,\Gamma^5\ri) -  \le(u \cdot \p_x  + i\, m\, v\ri) \le(\p_u \cdot \p_u - \p_v ^2 \ri) \ri] \Sigma(x,u,v)=0\,. \label{Fermionic CC 2}
\ee
where $v$ is an auxiliary variable corresponding to the 5th component of the auxiliary vector used to define the 5-dimensional generating functions.
Introducing the parameter $\m=ms$, the rescaled variables $\a=v/s$ and $\omega=u/\a$, and using the relation (\ref{dv du}), the equations (\ref{Bosonic CC 2}) and (\ref{Fermionic CC 2}) read
\bea
   \bigg[ & & (\p_\omega \cdot \p_x   - i\, m)  +  \frac{i\,\m}{s^2}~ (\omega \cdot \p_\omega)  \label{Bosonic CC 3} \\
   & & -\,\frac{1}{2} \le( \omega \cdot \p_x + i\, \m  \ri) \le( \p_\omega \cdot \p_\omega - \frac{1}{s^2} \le(\omega \cdot \p_\omega \ri)^2 + \frac{2 (s-1)}{s^2} ~ \omega \cdot \p_\omega - \frac{s-1}{s}\ri)  \bigg] J(x,\omega)=0\,, \nonumber
\eea
\bea
   \bigg[ & & \le(\Gamma \cdot \p_\omega - i\, \Gamma^5 + \frac{i}{s}~ \Gamma^5 ~ \omega \cdot \p_\omega \ri) \le( \Gamma \cdot \p_x + m \Gamma^5 \ri)    \label{Fermionic CC 3} \\
   & & -\, \le( \omega \cdot \p_x + i\, \m  \ri) \le( \p_\omega \cdot \p_\omega - \frac{1}{s^2} \le(\omega \cdot \p_\omega \ri)^2 + \frac{2 (s-1)}{s^2} ~ \omega \cdot \p_\omega - \frac{s-1}{s}\ri)  \bigg] \boldsymbol{\sigma}(x,\omega)=0\,,  \nonumber
\eea
where $J(x,\omega)$ and $\boldsymbol{\sigma}(x,\omega)$ are the new currents defined by
\be
j(x,u,v)=\a^s \, J(x,\omega)\,, \quad \quad  \Sigma(x,u,v)=\a^s \,\boldsymbol{\sigma}(x,\omega)\,.
\ee
Taking the Khan-Ramond limit \cite{KR} $(m\rightarrow 0$, $s \rightarrow \infty$, $ms=\m)$ of equations (\ref{Bosonic CC 3}) and (\ref{Fermionic CC 3}), we find
\be
   \le[ \p_\omega \cdot \p_x  -\,\tfrac{1}{2} \le( \omega \cdot \p_x + i\, \sigma \m  \ri) \le( \p_\omega \cdot \p_\omega - \sigma\, \ri)  \ri] J(x,\omega)=0\,, \label{Bosonic current 1}
\ee
\be
   \le[ \le(\gamma \cdot \p_\omega + \sigma\, \ri) \le( \gamma \cdot \p_x \ri) -\, \le( \omega \cdot \p_x + i\, \sigma \m  \ri) \le( \p_\omega \cdot \p_\omega - \sigma\, \ri)  \ri] \boldsymbol{\sigma}(x,\omega)=0\,, \label{Fermionic current 1}
\ee
with $\sigma=1$, and where $\gamma^\m= i \,\Gamma^5 \Gamma^\m$ are new gamma matrices which satisfy the Clifford algebra. Writing the parameter $\sigma$
\footnote{We choose bold $\boldsymbol{\sigma}$ for the fermionic current, and tiny $\sigma (=0$ or $1)$ for the parameter distinguishing HSP and CSP theories.}
in (\ref{Bosonic current 1}) and (\ref{Fermionic current 1}) we also cover the higher-spin conservation conditions when $\sigma=0$ (Eqs. (\ref{Bosonic CC 1'}) and (\ref{Fermionic CC 1'})).
Applying dimensional reduction on (\ref{B F constraints}) and following the above way, we find the constraints on the currents
\be
 \le(\p_\omega \cdot \p_\omega - \sigma\,  \ri)^2 \,  J(x,\omega)=0\,, \quad \quad \le( \gamma \cdot \p_\omega -\sigma\, \ri) \le( \p_\omega \cdot \p_\omega - \sigma\, \ri) \,\boldsymbol{\sigma}(x,\omega)=0\,. \label{B F constraint 2}
\ee

To find the interactions with a background current presented in \cite{ST PRD, BNS}, we should work in $\e$-space, as in the previous sections. The constraints of (\ref{B F constraint 2}) in $\e$-space lead to
\be
\widetilde{J}(x,\e)=\delta'(\e^2+\sigma\,) J(x,\e) \,, \quad \quad \widetilde{\boldsymbol{\sigma}}(x,\e)=\delta'(\e^2+\sigma\,)\, (\gamma \cdot \e -i \,\sigma\,) \boldsymbol{\sigma}(x,\e)\,, \label{J & sigma}
\ee
where $J(x,\e)$ and $\boldsymbol{\sigma}(x,\e)$ are unconstrained arbitrary functions. If we write (\ref{Bosonic current 1}) and (\ref{Fermionic current 1}) in $\e$-space and use (\ref{J & sigma}), we find
\be
\delta(\e^2+\sigma\,) \le( \p_\e \cdot \p_x + \sigma \m \ri)  J(x,\e) =0 \,,  \label{Bosonic J}
\ee
\be
 \delta(\e^2+\sigma\,) \le( \gamma \cdot \e - i\, \sigma\, \ri) \le( \p_\e \cdot \p_x +\sigma \m \ri)  \boldsymbol{\sigma}(x,\e) =0 \,.
\ee
These conservation conditions can be obtained from the invariance of the following interaction terms
\be
S_{int}= -\,{ \int  d^4 x\, d^4\eta }\,\,  \delta'(\e^2+\sigma\,) \, J(x,\e)\, \Phi(x,\e)\,, \label{B S_int}
\ee
\be
S_{int}  =  i{ \int  d^4 x\, d^4\eta }\,\,
\delta'(\eta^2+\sigma\,) \le[\, \overline{{\boldsymbol{\sigma}}}(x,\eta) (\gamma \cdot \eta + i\,\sigma\,)  \Psi(x,\eta)  - \,
\overline{\Psi} (x,\eta) (\gamma \cdot \eta - i \, \sigma\,) \boldsymbol{\sigma}(x,\eta) \ri], \label{F S_int}
\ee
under the gauge transformations (\ref{Bosonic GT eta 1}) and (\ref{Fermionic GT eta}). The currents and gauge fields are now both unconstrained.

\section{Minimal coupling between scalar matter and continuous-spin gauge fields}\label{s6}

Minimal interactions of higher-spin gauge fields with scalar matter and the corresponding long-range interactions between two scalars, mediated by massless higher-spin particles propagating on Minkowski background, were investigated in \cite{BJM}, by making use of the Berends-Burgers-van Dam (BBvD) currents \cite{BBvD} which are symmetric conserved currents of rank $s$, bilinear in the scalar field and containing $s$ derivatives. More precisely, following Noether's method, and introducing cubic couplings between a complex scalar field and a tower of symmetric tensor gauge fields, the amplitude for the elastic scattering of four scalars was found in any dimension (see also \cite{Taronna:2011kt,Ponomarev:2016jqk,Sleight:2016xqq} for more recent studies of such a process).
In this section, following the procedure of \cite{BJM}, we introduce a BBvD-like current of rank $s$, which is bilinear in two real scalar fields and includes $s$ derivatives, obeying a conservation-like law using the equations of motion. We find a generating function of these conserved-like currents and present Noether interactions between the scalar and gauge fields.


In order to find an appropriate background current contracting with continuous-spin fields, we introduce the symmetric conserved-like currents (BBvD-like currents) of rank $s$ as \footnote{The conserved-like currents, with equal scalar fields $\phi_1=\phi_2$\,, are proportional to the conserved currents already introduced in \cite{BJM, BBvD} up to a dimensionful parameter\,.}
\be
\sd {\mathcal{J}}{\mu_1\ldots\mu_s}(x)\, = \, ~\lambda^{( D - 6 + 2 s ) / 2}~ ~ \phi_{2}(x) ~ \overset\leftrightarrow{ \p}_{\m_1} \cdots  \, \overset\leftrightarrow {\p}_{\m_s}  ~\phi_{1}(x)  \,,   \label{BBvD 1}
\ee
where $D$ is the spacetime dimension, $\lambda$ is a parameter with the dimension of a length\,\footnote{The parameter $\lambda$ is added so that the tensor currents $\mathcal{J}^{(s)}$ have mass dimension $(D+2)/2$\,. }, $\phi_1$ and $\phi_2$ are real scalar fields of masses $m_1$ and $m_2 $\, and the double arrows are defined by
 \be
 A(x) \, \overset\leftrightarrow{ \p}_x  \, B(x) \, := \,  A(x) \, \overset\rightarrow{ \p}_x \, B(x) \, - \, B(x) \, \overset\leftarrow{ \p}_x \, A(x)\,.
 \ee

Consider a matter action made of two free scalar fields $\phi_1$ and $\phi_2$ of masses $m_1\geqslant 0$ and $m_2\geqslant 0$,
\be
S_0[\,\phi_i\,]~=~ \tfrac{1}{2} \sum_{i =1,2} \, \int \, d^D x    \, \le(~  \, \e^{\m\n}~ \p_\m \phi_i(x) ~ \p_\n \phi_i(x)\, -\,  m_i ^2 ~ [\,\phi_i(x)\,]^{\,2} ~\ri)
\ee
which leads to the Euler-Lagrange equations $\le(\Box_x + m_1 ^2\ri) \phi_1(x)=0$ and $\le(\Box_x + m_2 ^2\ri) \phi_2(x)=0$.
The currents \eqref{BBvD 1} are unconstrained (traceful)
\be
\e^{\m_1 \m_2}~\sd {\mathcal{J}}{\mu_1\ldots\mu_s}(x)~ {\approx \hspace{-10pt}\slash\hspace{0pt}} ~ 0 \,,
\ee
even when the scalar fields are massless and obey a conservation-like law
\be
\p^{\,\nu} ~ \sd {\mathcal{J}}{\nu\mu_1\ldots\,\mu_{s-1}}(x) ~ + ~ {\mathbf{\m}} ~  \overset{(s-1)}{\mathcal{J}}_{\mu_1\ldots\,\mu_{s-1}}(x) ~ \approx ~ 0\,,
\label{Continuity-like}
\ee
where the ``weak equality'' symbol $\approx$ means we used the equations of motion for the free scalar fields and
\be\label{mudiffm}
\m = \lambda \,\Big((m_1)^2 - (m_2)^2\Big)
\ee
is a positive parameter with the dimension of a mass\,\footnote{For equal masses $m_1=m_2$\,, the BBvD-like currents ${\mathcal{J}}^{(s)}$ become the BBvD ones \cite{BBvD}\,.} which can be identified with the parameter characterising the continuous-spin representation if the difference of mass squares is fine tuned, as will be assumed from now on.
These currents can be packed into a generating function\footnote{\,
 Since $\e_\m$ is dimensionless, from \eqref{BBvD 1}, it is clear that the generating function $\mathcal{ J}(x,\e) $ has mass dimension $(D+2)/2$\,.}
\be
\mathcal{ J}(x,\e) \,=\, \sum\limits_{s = 0}^{\infty}\, \frac{~1~}{s!} ~ \e^{\mu_1}\ldots \e^{\mu_s} ~ \sd {\mathcal{J}}{\mu_1\ldots\mu_s}(x)=\lambda^{(D-6)/2}~ \phi_2 (\,x - \lambda\,\e\, ) ~ \phi_1 (\,x +\lambda\,\e\,)\,, \label{BBvD}
\ee
and we remind the reader that $\e$ is a dimensionless auxiliary vector.
This function obeys
\bea
(\p_\e \cdot \p_x)\mathcal{ J} (x,\e) \,  &= & ~\lambda^{(D-4)/2}~ \Big[ \,  \phi_2 (X_- ) \, \Box_{X_+} \, \phi_1 (X_+)
\, -\, \phi_1 (X_+) \, \Box_{X_-}\, \phi_2 (X_- ) \Big] \nonumber \\
& \approx & - \,\mu \, \mathcal{ J}(x,\e)\, ,  \label{current 2}
\eea
where $X_{\pm} \,:=\, x \, \pm \,\lambda\,\e $.
Thus we can write the conservation-like condition \eqref{Continuity-like} in terms of the generating function as
\be
\le( \p_\e \cdot \p_x + \,\m \, \ri)  \mathcal{J}(x,\e) \, \approx \, 0\,. \label{Int 5}
\ee
This exhibits that the tower of BBvD-like currents \eqref{BBvD 1} can be used as a suitable choice of current to couple to a continuous-spin gauge field (\ref{B S_int})\,.
Indeed, consider the conservation-like condition (\ref{Bosonic J}) with $\sigma=1$\,, and rewrite it as
\be\label{convlike}
\le( \p_\e \cdot \p_x +  \m \ri)  J(x,\e) = \le(\e^2 + 1 \ri) \, \a(x,\e)\,,
\ee
with $\a(x,\e)$ an arbitrary function. One finds that $J=\mathcal{J}$ is a solution  with $\alpha=0$ of the conservation-like equation \eqref{convlike} on the mass-shell of the two scalar fields.

The conclusion is that two distinct scalars with distinct masses such that \eqref{mudiffm} can interact minimally with a CSP via the infinite collection of BBvD currents \eqref{BBvD 1}.


\section{Current exchange}\label{s7}

The response of unconstrained higher-spin gauge fields to external currents was investigated systematically in \cite{FMS}\,.
In this section, we shall obtain the current exchange from the Schuster-Toro action for
unconstrained bosonic CSPs. 
On the one hand, the Euclidean analytic continuation (proposed in \cite{ST PRD} as a regulated version) of this action produces the expected residue of the propagator and 
reproduces the result for the local unconstrained HSPs,
as will be explain in Subsection \ref{Euclid}. On the other hand, as we will discuss in Subsection \ref{propag}, the actual Lorentzian action appears to propagate unphysical degrees of freedom.
This result indicates that the action, although it correctly describes the free dynamics,
 is not a good starting point to add interactions. A similar situation was encountered with higher spins in \cite{FMS}
where it was shown that a certain number of non-local actions resulting in
the correct free equations of motion do not reproduce the correct  current-current exchange and can thus
be excluded as starting point to introduce interactions.
The appendix \ref{eta_integrals} introduces the technical details linked to our discussions in this section.

\subsection{Propagated degrees of freedom}\label{propag}

From the bosonic action with source, i.e. the sum of \eqref{Bosonic Action} and \eqref{B S_int}, one obtains the equation of motion
\bea
&&\d'(\e^2 +\sigma)\le\{\,\le[ - \, \Box_x  + (\e \cdot \p_x) \, \left( \p_{\e} \cdot \p_x + \sigma \m \right) -  \tfrac{1}{2} (\e^2 + \sigma) \left( \p_{\e} \cdot \p_x + \sigma \m \right) ^2 \ri] \Phi(x,\e)\, -\,  J(x,\e) \,\ri\}\nonumber\\
&&\qquad\qquad\qquad\qquad\qquad\qquad\qquad\qquad\qquad\qquad\qquad\qquad\qquad\qquad = 0\,,  \label{EOM  J}
 \eea
where $\Phi$ and $J$ are the generating functions in $\e$-space. Considering the $\chi$ symmetry \eqref{kapa transformation}\,, we can write \eqref{EOM  J} as
 \be\label{J+pg}
 \le[ - \, \Box_x  + (\e \cdot \p_x) \, \left( \p_{\e} \cdot \p_x + \sigma \m \right) -  \tfrac{1}{2} (\e^2 + \sigma) \left( \p_{\e} \cdot \p_x + \sigma \m \right) ^2 \ri]  \Phi
 \,=\, J\,+\,\mbox{pure gauge}\,,
 \ee
where by ``pure gauge'' one means that the right-hand-side in \eqref{J+pg} takes the form of the right-hand-side in \eqref{kapa transformation}.
By taking into account $\epsilon$ symmetry \eqref{Bosonic GT eta 1}\,, one can even write
 \be
  - \, \Box_x \, \Phi(x,\e) \, = \,  J(x,\e)  \, +\,  \mbox{pure gauge} \,,
 \ee
where now ``pure gauge'' means a right-hand-side of the form \eqref{Bosonic GT eta 1} with $\epsilon=\left( \p_{\e} \cdot \p_x + \sigma \m \right)\Phi$\,.
 Therefore, we can solve the bosonic equation of motion with source \eqref{EOM  J} as
 \be\label{phij}
 \Phi(x,\e) ~ = ~ - \, \frac{1}{\,\,\, \Box_x } \, \,  J(x,\e) \,+\,\mbox{pure gauge} \,.
 \ee
The current exchange is proportional to the interaction term \eqref{B S_int} evaluated on the solution \eqref{phij}, i.e.
\be
S_{int}~=~{ \int  d^D x\, d^D\eta }\,\,  \delta'(\e^2+\sigma\,) \, J(x,\e) ~ \frac{1}{\,\,\, \Box_x } ~  J(x,\e) \,,  \label{chii}
\ee
where the ``pure gauge'' contribution disappeared due to the conservation-like condition.

The generating function $J(x,\e)$ obeys the conservation-like condition \eqref{convlike}, therefore the Fourier transform $J(p,\e)$ of the generating function can be assumed to obey
\be
(p\cdot\partial_\e + i\,\mu)J(p,\e)=0\,.
\ee
The current exchange \eqref{chii} in momentum space reads
\be
S_{int}~=~{ -\int  d^D p\, d^D\eta }\,\,  \delta'(\e^2+\sigma\,) \, J(-p,\e) ~ \frac{1}{\,\,\, p^2} ~  J(p,\e) \,,  \label{chiii}
\ee
Using light-cone coordinates \eqref{Null} adapted to a given mode, i.e. such that $p^+\neq 0$ is the only non-vanishing component, the conservation-like condition reads $(p^+\partial_{\e^+} \,+\, i\,\mu)J(p,\e)=0$, which can be solved as $J(p,\e)=\exp(-\,i\,\frac{\mu}{p^+}\,\eta^+)\,j(p,\eta)$ where $\tfrac{\p}{\p \e^+}\, j(p,\e) =0\,$. Therefore, the $\eta^+$ dependence drops from the integrand in \eqref{chiii}.

In order to identify the propagating degrees of freedom without making use of the delicate Wick rotation to the Euclidean signature, let us denote the remaining integrand as
\be
f(p,\e)\,:=\,j(-p,\e) \,\frac{1}{~p^2}\, j(p,\e)
\ee
in the current exchange \eqref{chii} and consider the derivative of the Dirac delta distribution as defining the measure of the $\e$-integral.

From now, we will leave the momentum dependence implicit  and will focus on the $\e$-integral part of the current exchange, which can be written generically as
\bea
I\,&=& \, \int d^D \e ~ \delta'(\e^2 + \sigma ) ~ f(\e) \,, \\
&=& \frac{d}{d\sigma}~  \int d^D \e ~ \delta(\e^2 + \sigma ) ~ f(\e) \,,
\eea
where the function $f(\e)$ is actually independent of $\eta^+$.
By changing the variable $\e $ to $\e=\sqrt{\sigma}\,  \tilde{\e}$\,, and then taking the derivative with respect to $\sigma$\,, we get
\bea
I\,&=&\, \le(\tfrac{D}{2}-1\ri) ~ \sigma^{\frac{D}{2}-2} ~ \int d^D \tilde{\e} ~ \delta(\tilde{\e}^2 + 1 ) ~ f(\sqrt{\sigma}\, \tilde{\e} )  \\
& &\quad\quad~~ +~ \sigma^{\frac{D}{2}-1}~ \int d^D \tilde{\e}  ~ \delta(\tilde{\e}^2 + 1 ) ~ \frac{\tilde{\e}^{\,\mu}}{2\, \sqrt{\sigma}} ~
 \frac{\p}{\p \e^\mu
} ~ \Big[\,f(\sqrt{\sigma}\, \tilde{\e} )\,\Big]  \\
 &=& \sigma^{\frac{D}{2}-2}~ \int d^D \tilde{\e} ~ \delta(\tilde{\e}^2 + 1 ) ~ \le[ \le(\tfrac{D}{2}-1\ri)
  + \frac{1}{2} \, \tilde{\e}^{\,\mu} ~  \frac{\p}{\p \tilde{\e}^{\,\mu} }    \ri] ~ f(\sqrt{\sigma}\, \tilde{\e} )  \,.
\eea
If we put $\sigma=1$, choose light-cone coordinates \eqref{Null}, make use of $\tfrac{\p}{\p \e^+}\, f (\e) =0$, then the latter integral can be written as
\be
I \, = \, \frac{1}{2} \,  \int d^D \e ~ \delta(\e^2 + 1 ) ~ \le[ \frac{\p}{\p \e^i} \, \le(\e^i \, f(\e) \ri) \,+\, \e^- \, \frac{\p}{\p \e^-} \, f(\e)   \ri]\,,
\label{Null-cone}
\ee
where the index $i$ corresponds to the $D$-2 transverse coordinates.
In these coordinates \eqref{Null}, the Dirac delta function in \eqref{Null-cone} takes the form
\bea
\delta(\e^2 +1)\, &=& \, \delta\le(2\, \e^+\e^- - \e^i \e_i +1 \ri) \\
&=&{ \lim_{\epsilon\to 0}} ~ \le\{ \tfrac{1}{~2 \, \sqrt{ (\e^-)^2 + \epsilon^2 \,}~} ~ \delta \le( \e^+ - \tfrac{sign(\e^-)}{~2 \, \sqrt{ (\e^-)^2 + \epsilon^2 \,}~}\,( \e^i \e_i -1 )\, \ri) \ri\} \,,
\label{delta}
\eea
where we introduced the regularisation
\be
\e^- \,=\, \lim_{\epsilon\to 0} \le\{\, sign(\e^-) ~\sqrt{ (\e^-)^2 + \epsilon^2 \,}  \ri\}
\ee
in order to extract a factor $\eta^-$ from the delta function (since the above expression is not valid for $\eta^-=0$). Now, by applying \eqref{delta} for \eqref{Null-cone} and performing the integral with respect to $\e^+$, we get
\bea
I\,&=&\,\lim_{\epsilon\to 0}~   \int d\e^- \, d^{D-2}\e^i ~~ \frac{1}{~4\sqrt{ (\e^-)^2 + \epsilon^2 \,}~} ~ \le[ \frac{\p}{\p \e^i} \, \le(\e^i \, f(\e) \ri) \,+\, \e^- \, \frac{\p}{\p \e^-} \Big(\,f(\e)\,\Big)   \ri] \\
&=& \,\lim_{\epsilon\to 0}~   \int d\e^- \, d^{D-2}\e^i ~~ \frac{\e^-}{~4\sqrt{ (\e^-)^2 + \epsilon^2 \,}~}  ~ \frac{\p}{\p \e^-} \,\Big( f(\e) \,\Big)\,,
\eea
since the integral of the divergence over the transverse coordinates vanishes. From
\bea
\lim_{\epsilon\to 0}~   \int d\e^- \, \frac{\e^-}{~\sqrt{ (\e^-)^2 + \epsilon^2 \,}~}  ~ \frac{\p}{\p \e^-} \,\Big( f(\e)\,\Big) \,&=&\,
 \int_0^{\infty} d\e^- ~  \frac{\p}{\p \e^-} \, f(\e) -  \int^0_{-\infty} d\e^- ~  \frac{\p}{\p \e^-} \,\Big( f(\e)\,\Big) \nonumber \\
 &=& -\, 2 \, f\le(\e^- =0 , \e^i\ri)  \,,
\eea
we finally find
\be
I\,=\, -\, \frac{1}{2} \, \int \, d^{D-2}\e^i  ~~ f\le(\e^- =0 , \e^i\ri) \,.
\ee
As one can see, the integral is not localised on the hypersphere $\e^i\e_i=\mu^2$. Therefore, the current exchange \eqref{chiii} in the physical (i.e. Lorentzian) signature appears to propagate a continuum of CSPs ($\e^i\e_i>0$) rather than a single CSP ($\e^i\e_i=\mu^2$).

\subsection{Euclidean analytic continuation}\label{Euclid}

Let us now turn to the Euclidean analytic continuation of the current-exchange \eqref{chii}. For notational simplicity, we will not distinguish the Wick-rotated variables. Similarly, the wave operator will actually stand for (minus) the Laplacian.

Referring to appendix \ref{eta_integrals}\,, by using \eqref{Int 4-4} we can rewrite the Wick rotation of the current-exchange \eqref{chii} as
\be
 S_{int} = \sigma^ {\frac{D}{2} - 2}\int  d^D x \sum_{k=0} ^{\infty}  A_{(\frac{D-4}{2}, \,k)}
   \le[ {\cal J}{_{\left(\frac{D-4}{2},\,k\right)}} (x,\e\,;\sigma) \ri]  \frac{ \overset{(k)}{\mathbb{P}}_{\frac{D-4}{2}}\, (\overset\leftarrow{\p}_\e, \overset\rightarrow{\p}_\e )}{\Box_x}
    \le[ {\cal J}{_{\left(\frac{D-4}{2},\,k\right)}} (x,\e\,;\sigma) \ri] \Bigg| _ {\e \,=\, 0} \,,  \label{current exchange}
 \ee
where the coefficients $A_{(\nu, k)}$ and the operator $\mathbb{P}^{(k)}_\nu$, which we will call ``spin-$k$ current-exchange operator'', are defined respectively in \eqref{A} and \eqref{Current operator 2}. We also made use of the redefined conserved(-like) currents
\be
{\cal J}{_{\left(\frac{D-4}{2},\,k\right)}} (x,\e\,;\sigma)  =
\sigma^\frac{k}2\,\, _0 \,F_1 \, \Big( \,\tfrac{D-2}{2} + k\, ; \, -\, \sigma \, \frac{\,  {\p}_\e \cdot {\p}_\e}{4}   \, \Big)\,J(x\,,\sigma^{-\frac12}\,\eta)\,,
\ee
following  \eqref{redefined fields}.

The term at given $k=s$ reproduces the current-exchange for the local unconstrained spin-$s$ field (we set $\sigma=1$ for simplicity)
\be\label{calJ}
{\cal J} ~ {\overset{(s)}{\mathbb{P}_{\frac{D-4}{2}}}\, (\overset\leftarrow{\p}_\e, \overset\rightarrow{\p}_\e )} ~ {\cal J} \, \Big| _ {\e \,=\, 0} =
\sum_{n=0} ^{\lfloor s/2 \rfloor} \frac{1}{\,2^{2 n}\, (\,3-\tfrac{D}{2}-s\,)_n } \,
\, \frac{s!}{ n! \, (s - 2 n)!} ~{\overset{(s)}{ {\cal J} }}{}^{[\,n\,]} \cdot {\overset{(s)}{ {\cal J} }}{}^{[\,n\,]} \,,
\ee
where
\be
{\cal J}(x,\e;1) = \sum_{s=0}^{\infty} \, \frac{1}{s!} ~ \e^{\m_{1}}   \ldots \e^{\m_{s}} ~\overset{(s)}{{\cal J}} _{\,{\m_{1}\ldots\m_{s}}}(x)
 \,,
\ee
and
the superscript ${[\,n\,]}$ in \eqref{calJ} denotes the n\emph{th} trace of the rank-$s$ conserved current ${\cal J}^{(s)}$, morover this latter object has $s-2n$ contractions, denoted by a dot, with another one. The expression \eqref{calJ} is exactly the current exchange obtained by the authors of \cite{FMS}\,.

\section{Conclusions and future directions}\label{s8}

We have explained how to obtain the bosonic and fermionic Segal-like actions of \cite{ST PRD, BNS} respectively from the Fronsdal-like and Fang-Fronsdal-like equations of \cite{BM} by solving the trace constraints in Fourier space.

We have also derived the conservation-like conditions which must be obeyed by external sources added in the right-hand-side of the (Fang-)Fronsdal-like equations from \cite{BM}. By solving trace constraints again, one can then rederive the conservation-like condition of \cite{ST PRD} that must be obeyed by a current to which the bosonic continuous-spin field can couple minimally. Accordingly, we introduced a current (similar to the one of Berends, Burgers and Van Dam \cite{BBvD}) which is bilinear in a pair of scalar matter fields, satisfying this condition when the difference of their mass squared is properly tuned.
It would be interesting to compare this current with the stress-energy momentum tensor obtained from the Khan-Ramond infinite-spin massless limit in \cite{Rehren:2017xzn}.

Finally, we investigated the current exchange in the formulation of \cite{ST PRD}. In Euclidean signature, the propagator extracted from the Schuster-Toro action has been shown to be related to the one of a tower of (redefined) fields of finite/discrete-spin.
However, in Lorentzian signature the propagator extracted from the Schuster-Toro action does not appear to propagate the right degrees of freedom, in the sense that it appears to propagate a continuum of CSPs rather than a single one. In this respect, the Metsaev actions appear to be better candidates for defining suitable propagators, given that they are Fronsdal-like actions. Nevertheless, the fact that the Segal-like action formally defines the correct current exchange in \textit{Euclidean} signature suggest that there might exist either a slight improvement of the action or a suitable prescription to obtain the correct propagator in \textit{Lorentzian} signature. We leave these open issues for further exploration.

\paragraph{Note added:} While this work was in preparation, the paper \cite{Metsaev:2017cuz} appeared which contains an extensive discussion of the consistent cubic vertices for bosonic continuous-spin fields and massive fields of arbitrary spin.

\acknowledgments

We thank Dario Francia, Massimo Taronna and Evgeny Skvortsov for useful discussions. X.B. and M.N. are especially grateful to Mohammad Khorrami for several technical tips on special functions. M.N. is also grateful to Mohammad Mehdi Sheikh-jabbari and Hamid Reza Afshar for their support, and to the LMPT and IPM members for their kind hospitality.

The research of X.B. was supported by the Russian Science Foundation grant 14-42-00047 in association with the Lebedev Physical Institute.

\appendix

\section{Conventions}\label{appA}

Our gamma matrices in 4 dimensions satisfy the Clifford algebra $\le\{\gamma^\m,\gamma^\n\ri\}=2\,\e^{\m\n}$\,.
We use the mostly negative convention for the spacetime signature. Therefore,
\be
\le(\gamma^0\ri)^\dag= +\, \gamma^0\,, \quad \le(\gamma^i\ri)^\dag= - \,\gamma^i\,, \quad \le(\gamma^\m\ri)^\dag = \gamma^0 \gamma^\m \gamma^0\,, \quad
\le(\gamma^5\ri)^\dag= +\, \gamma^5\,,
\ee
where $\gamma^5:=i\,\gamma^0 \gamma^1 \gamma^2 \gamma^3$ anticommutes with all $\gamma^\m$s.

In 5-dimensional spacetime, gamma matrices are still 4$\times$4 matrices and are defined as
\be
\gamma^M = \le\{ \gamma^0, \gamma^1, \gamma^2, \gamma^3, \gamma^4=\mp\, i \gamma^5 \ri\}\,, \quad M=0,1,2,3,4 \,,
\ee
which satisfy the Clifford algebra of 5-dimensional spacetime. In odd dimensions, we have two independent representations depending on the choice of sign for $\gamma^4$. Among these two possibilities, we chose the representation with minus sign in $\gamma^4$ in the body of paper.

The light-cone coordinates are defined as:
\be
\e^\pm\,=\, \frac{1}{\sqrt{2}} \le( \e^0 \pm \e^{D-1} \ri)\,,  \label{Null}
\ee
and
\be
\e^i\,=\, \le( \,\e^1, \ldots , \e^{D-2}\,  \ri)\,,
\ee
thus $\e^2 = 2\,\e^+ \e^- - \e^i \e_i$\,.

\section{Fermionic continuous-spin equations}\label{Fermionic CSP equation}

In this Appendix, we review the procedure of \cite{BM} to obtain a Fang-Fronsdal-like equation from the massive higher-spin equation in more details. We also review the equation of motion from \cite{BNS} which can be obtained from an action principle and which is equivalent to the previous Fang-Fronsdal-like equation.

\subsection{Massless equation}

We consider the Fang-Fronsdal equation in 4 dimensions \cite{BM}
\be
\Big[(\gamma\cdot \partial_x)  - (u\cdot\partial_x)
\left(\gamma\cdot{\partial_{u}}\right)\Big]\Psi (x,u)=0\,,  \label{HS 1}
\ee
where $\Psi$ is a 4-component massless totally symmetric spinor-tensor introduced in (\ref{psi}).
This equation is gauge invariant under the gauge transformation
\be
\delta \Psi=( u\cdot \partial_x )\,\epsilon\,,  \label{HS 2}
\ee
with the gamma-trace condition
\be
\left(\gamma\cdot \partial_u\right)\,\epsilon=0\,,  \label{HS 3}
\ee
where $\epsilon$ introduced in (\ref{epsilon}).
The triple gamma-trace condition reads
\be
\left(\gamma\cdot\partial_u\right)
\left(\partial_u\cdot \partial_u\right)\Psi=0\,.  \label{HS 4}
\ee

\subsection{Massive equations}

To obtain the massive fermionic equations, we take into account the massless ones in 5 dimensions and consider a mode with $p^4=m$. The equation of motion (\ref{HS 1}) becomes
\be
\Big[(\gamma\cdot \p_x  + i m \gamma^4)
 - \le(u\cdot \p_x +i m v\ri) \le( \gamma\cdot{\partial_{u}}+ \gamma^4 {\partial_{v}} \ri)\Big]\Psi (x,u,v)=0,
\ee
where we put $\partial_{x_4}=i m$, $u^4=v$, and $\gamma^4 =\mp\,i\gamma^5$ is the fifth element of the Clifford algebra in $5$ dimensions.
Therefore, we have two possible independent equation of motions
\be
\Big[(\gamma\cdot \p_x  \pm m \gamma^5)
 - \le(u\cdot \p_x +i m v\ri) \le(z \gamma\cdot{\partial_{u}} \mp \, i\gamma^5 {\partial_{v}} \ri)\Big]\Psi_{\pm} (x,u,v)=0\,, \label{HS I}
\ee
as are expected in odd dimensions depending on the choice of sign. The upper and lower signs are related to the massive fermionic fields $\Psi_{+}$ and $\Psi_{-}$ respectively.
These equations are invariant under the gauge transformations
\be
\delta \Psi_{\pm}=\big(u \cdot \p_x + i m v\big)\,\epsilon_{\pm}\,,  \label{HS II}
\ee
with the gamma-trace conditions
\be
\big( \gamma\cdot \partial_{u} \mp i \gamma^5 \partial_{v} \big) \,\epsilon _{\pm}=0\,. \label{HS III}
\ee
The triple gamma-trace conditions become,
\be
\big( \gamma\cdot\partial_{u}\mp\, i \gamma^5 \partial_{v}\big)
\big( \partial_{u}\cdot \partial_{u} - {\partial^2_{v}}\, \big)  \Psi_{\pm}=0\,. \label{HS IV}
\ee

In order to obtain the massless equations from the massive ones by taking a limit, we introduce the parameter $\m$ and the variable $\a$
\be
\m=m\, s , \quad \quad \a=\frac{v}{s},
\ee
and demand when $s$ goes to infinity and mass to zero, that $\m$ and $\a$ are kept fixed. Defining the new gauge fields $\psi_{\pm}$ and gauge parameters $\varepsilon_{\pm}$  by
\be
\Psi_{\pm}(x,u,v)= \a^s \,\psi_{\pm}(x,\omega) \, ,\quad  \quad \epsilon_{\pm}(x,u,v)= \a^{s-1} \,\varepsilon_{\pm} (x,\omega) \,,
\ee
where $\omega:=u/\a$, and using the following relations
\be
{\partial_u}={\a^{-1}}\,{\partial_\omega}\,,\quad\quad
{\partial_v}={1\over s} \,{\a^{-1}} \,  \Big(-\, \omega \cdot {\partial_\omega}+{\a \, \partial_\a}\Big)\,, \label{dv du}
\ee
we will be able to rewrite the equations (\ref{HS I})$-$(\ref{HS IV}) respectively as
\be
\le[ (\gamma\cdot \p_x  \pm m \gamma^5)
 - \le(\omega\cdot \p_x +i \mu \ri)  \le(  \gamma\cdot{\partial_{\omega}} \mp \, i\gamma^5   \le[- \, \tfrac{1}{s} \, \omega \cdot{\partial_{\omega}} +1 \ri] \ri)  \ri] \psi_{\pm} (x,\omega)=0\,, \label{HS I 1}
\ee
\be
\delta \psi_{\pm}=\le(\omega \cdot  \p_x+ i \mu\ri)\,\varepsilon_{\pm}\,,  \label{HS II 2}
\ee
\be
\le( \gamma\cdot \partial_{\omega} \mp i \gamma^5  \le[- \, \tfrac{1}{s} \, \omega \cdot{\partial_{\omega}} + \tfrac{s-1}{s}  \ri] \ri) \varepsilon_{\pm}=0\,,
\label{HS III 3}
\ee
\be
\le( \gamma \cdot \partial_{\omega}\mp\, i \gamma^5  \le[-\, \tfrac{1}{s} \, \omega \cdot \p_\omega + \tfrac{s-2}{s} \ri ]  \ri)
  \le( \p_\omega \cdot \p_\omega - \tfrac{1}{s^2} \le(\omega \cdot \p_\omega \ri)^2 + \tfrac{2 (s-1)}{s^2} ~ \omega \cdot \p_\omega - \tfrac{s-1}{s}\ri)  \psi_{\pm}=0\,. \label{HS IV 4}
\ee
These relations are equations for massive HSP.

\subsection{Continuous-spin equations}

Taking the limit (\ref{Limit}), the equations (\ref{HS I 1})-(\ref{HS IV 4}) will reproduce the Fang-Fronsdal-like equations of \cite{BM}.
In order to obtain equations in the form presented in \eqref{Fermionic}-\eqref{Fermionic F constraint}, we have to  multiply the limit (\ref{Limit}) of (\ref{HS I 1}) by $i \gamma^5$ to the left and make use of the new matrices $\C^\m = i \gamma^5 \gamma^\m$ which satisfy the same Clifford algebra as the $\gamma^\m$s.

In summary, the equations of motion for the fermionic CSPs become
\be
\le[ (\Gamma\cdot \p_x)
 - \le(\omega\cdot \p_x + i \m \ri)  \le(  \Gamma\cdot{\partial_{\omega}} \pm 1\ri)  \ri] \psi_{\pm} (x,\omega)=0\,, \label{HS I 1 1}
\ee
which are invariant under the gauge symmetries
\be
\delta \psi_{\pm}=\le(\omega \cdot  \p_x+ i \m\ri)\,\varepsilon_{\pm}\,,  \label{HS II 2 2}
\ee
with the gamma-trace conditions
\be
\le( \Gamma\cdot \partial_{\omega} \pm 1 \ri) \varepsilon_{\pm}=0\,.
\label{HS III 3 3}
\ee
The triple gamma-trace conditions read
\be
\le( \Gamma\cdot\partial_{\omega}\pm 1 \ri)
\le( \partial_{\omega}\cdot \partial_{\omega} - 1 \ri)  \psi_{\pm}=0\,. \label{HS IV 4 4}
\ee

\section{Integrals on auxiliary space}\label{eta integrals}

In this appendix, 
we express the existing $\e$-integrals in the actions as multiple contractions of indices whose structure can be nicely packed into special functions by making use of techniques similar to the ones presented in the appendix D of \cite{Segal:2002gd} (see also the appendix A of \cite{ST PRD}).

\subsection{Generating functions in Lorentzian signature}\label{generatingtitle}

The integrals over $\e$-space in the actions involve the Dirac delta function via $\delta (\e^2 +\sigma)$ or its derivative $\delta' (\e^2 +\sigma)$. We provide the general solution for the integrals containing $\delta^{(\ell)} (\e^2 +\sigma)$\,, where $\ell$ is the number of derivatives of the delta function with respect to its argument, but just $\ell=0,1$ are needed for this work.

In order to relate $\eta$-integrals to index contractions, we introduce the generating functions
\be
\texttt{G}^{(\ell)}(\omega\,;\sigma):= \int  d^D \e \,\, \delta^{(\ell)} (\e^2 +\sigma) \, e^{-i \,\e \cdot \omega}=\left(\frac{\partial}{\partial\sigma}\right)^\ell \texttt{G}^{(0)}(\omega\,;\sigma)\,,\label{Gsigmaa}
\ee
which are the Fourier transform of the corresponding distributions.
Since a function can be formally written as $F(\e)= \left[F(i \partial_\omega) \, {e^{{-i }{\eta \cdot \omega}} }\right]\big|_{\omega=0}$ we can write the $\e$-integrals in the action 
 as

\be
\int  d^D \e \, \delta^{(\ell)} (\e^2 +\sigma) \, F(\e) =
 \left[ F(i \,\partial_\omega) \, \texttt{G}^{(\ell)}(\omega\,;\sigma)\right]\bigg|_{\omega=0} =
 \left[ \texttt{G}^{(\ell)}(i \,\partial_\eta;\sigma) \, F(\eta)\right]\bigg|_{\eta=0}
  \,.   \label{Int F(etaa)}
\ee

For the Lorentzian signature, the integral \eqref{Gsigmaa} is effectively over an internal $(D-1)$-dimensional de Sitter space, more precisely on the one-sheeted hyperboloid $\e^2 =-\sigma$ in the $D$-dimensionsal Minkowski auxilliary space. This Lorentzian space has infinite volume, so these generating functions are not regular at the origin $\omega=0$, as will be seen from their explicit expression, but we will extract the regular parts.
One trick (c.f. \cite{Segal:2002gd,ST PRD}) to get a finite result is to consider the Euclidean signature, where the integral \eqref{Gsigmaa} is over the $(D-1)$-dimensional hypersphere.
We will see in the next subsection that the regular terms of \eqref{Gsigmaa}\,, for $\omega^2 < 0$\,, in the Lorentzian signature are exactly the ones obtained from \eqref{Gsigmaa} in the Euclidean signature with Wick-rotated coordinates.

\vspace{3mm}
Firstly, let us determine the explicit formulae for the generating functions \eqref{Gsigmaa} at $\ell=0$. They depend on the type of the auxiliary vector $\omega^\mu$. Let us consider the two important cases:
\vspace{1mm}

\noindent\textbf{Timelike case:} $\omega^2>0$

\noindent For $\omega^2>0$, one can perform the computation of the generating function \eqref{Gsigmaa} at $\ell=0$ by assuming without loss of generality that $\omega^0=\sqrt{\omega^2}\neq 0$ is the only nonvanishing component of the auxiliary vector $\omega^\mu\,$, and then write the final result in a manifestly covariant form.
The integral to perform is
\be
\texttt{G}^{(0)}(\omega\,;\sigma)= \int  d^D \e \,\, \delta (\e^2 +\sigma) \, e^{-i \,\e^0 \omega_0}=
\sigma^{\frac{D-2}2}\int d\eta^0 d^{D-1}\vec\eta\,\,\delta\big((\eta^0)^2-{\vec\eta}^{\,2}+1\big)\, e^{-i \,\sqrt{\sigma}\,\e^0 \omega_0}\,,
\ee
where $\vec\e=(\e^1,\cdots,\e^{D-1})$\,.
The integral over $\eta^0$ is performed via the identity
\be\label{deltax2}
\delta(x^2-a^2)=\frac1{2\,|a|}\left[\delta(x+a)+\delta(x-a)\right]\,,
\ee
which gives
\be
\texttt{G}^{(0)}(\omega\,;\sigma)=
\sigma^{\frac{D-2}2}\int_{{\vec\e}^{\,2}>1} d^{D-1}\vec\eta\,\,\frac1{\sqrt{{\vec\eta}^{\,2}-1}}\,\cos\big(\sqrt{\sigma({\vec\eta}^{\,2}-1)}\,\omega_0\big)\,.
\ee
Introducing the variable $t=\sqrt{{\vec\eta}^{\,2}-1}$ and performing the integral over the (D-2)-dimensional sphere corresponding to the remaining variables, one obtains that the generating function \eqref{Gsigmaa} reduces to the covariant expression
\be\label{diff2}
\texttt{G}^{(0)}(\omega\,;\sigma)= \sigma^{\frac{D-2}{2}}  ~ \frac{2 \, \pi^{\frac{D-1}{2}}}{~\Gamma\le(\frac{D-1}{2} \ri)~}
~ \int_{0}^{\infty} \, dt ~ (t^2 +1)^{\frac{D-3}{2}} ~ \cos (\,t \,\sqrt{\sigma \, \omega^2\,} \, ) \,.
\ee
This integral is defined as a distribution. For $D \leqslant 2 $, the integral is absolutely convergent; by using \eqref{Gamma} and \eqref{K}, reads
\be
\texttt{G}^{(0)}(\omega\,;\sigma)=2 \, \cos\le[ \pi \big( \tfrac{2 - D}{2} \big) \ri]
\left(\frac {\sqrt{\omega^2}}{ 2\pi\sqrt{\sigma}}    \right)^\frac{2-D}2\,K_{\frac{2-D}2}\left(\sqrt{\sigma\,\omega^2\,}\,\right)  \,,
\qquad (D \leqslant 2)
\label{Gsigma 0 K}
\ee
where $K$ is the modified Bessel functions of the second kind. 
%
Differentiating \eqref{diff2} twice with respect to $\sqrt{\omega^2}$ allows to obtain a recurrence relation for the generating functions $\texttt{G}^{(0)}$ of distinct $D$. This provides a way to compute $\texttt{G}^{(0)}$ for $D>2$ but we will refrain from doing so because it is the spacelike case that will be used in the sequel.
\vspace{1mm}

\noindent\textbf{Spacelike case:} $\omega^2<0$

\noindent For $\omega^2<0$\,, one can perform the computation of the generating function \eqref{Gsigmaa} at $\ell=0$ by choosing a frame where $\omega^0=0$ and then write the final result in a covariant way.
The integral to perform to compute the generating function \eqref{Gsigmaa} is
\be
\texttt{G}^{(0)}(\omega\,;\sigma)= \int  d^D \e \,\, \delta (\e^2 +\sigma) \, e^{i \,\vec\e\cdot\vec\omega}=
\sigma^{\frac{D-2}2}\int d\eta^0 d^{D-1}\vec\eta\,\,\delta\big((\eta^0)^2-{\vec\eta}^{\,2}+1\big)\, e^{i \,\sqrt{\sigma}\,\vec\e\cdot\vec\omega}\,,
\ee
One can perform the integral over $\e^0$ via \eqref{deltax2} which gives
\be
\texttt{G}^{(0)}(\omega\,;\sigma)=
\sigma^{\frac{D-2}2}\int_{{\vec\eta}^{\,2}>1} d^{D-1}\vec\eta\,\,\frac1{\sqrt{\vec\eta{}^{\,2}-1}}\, e^{i \,\sqrt{\sigma}\,\vec\e\cdot\vec\omega}\,.
\ee
Introducing spherical coordinates with the following two particular coordinates: the radial coordinate $\abs{\vec{\e}\,}:=\sqrt{\vec\eta{}^{\,2}}$ and the azimuthal coordinate $\theta$ defined by $\vec\e\cdot\vec\omega=\abs{\vec{\e}\,}\,\abs{\vec{\omega}}\,\cos\theta$\,, one can perform the integral over the (D-3)-dimensional hypersphere corresponding to the remaining coordinates
by making use of the integral representation of the Bessel function of the first kind \eqref{Bessel int}, we get
\be
\texttt{G}^{(0)}({\omega}\,;\sigma) \,=\, 2 \, \le( \pi\,\sqrt{\sigma} \,\ri) ^ {\frac{D-1}{2}}~
\le( \tfrac{2}{\,\,\abs{\vec{\omega}}\,} \ri)^ {\frac{D-3}{2}}~
\int_{1}^{\infty}\, d \abs{\vec{\e}} ~ \tfrac{\abs{\vec{\e}}^{\frac{D-1}{2}}  }{\sqrt{\, \abs{\vec{\e}}^2 -1}}
~ J_{\frac{D-3}{2}} \le( \sqrt{\sigma}\, \abs{\vec{\omega}} \, \abs{\vec{\e}}  \ri)\,.
\ee
This integral is divergent but we can multiply it by $\exp(-\,b \, (\,\abs{\vec{\e}}^2 -1 )^{1/2} )$, to be convergent for $D \geqslant 2 $\,. Then, by using \eqref{J integral} with $a=1$, $\nu=(D-3)/2$, $y=\sqrt{\sigma}|\vec\omega|$, and finally taking the limit $b\rightarrow 0$, we will get\footnote{We thank Mohammad Khorrami for his useful comments on this point.}
\be
\texttt{G}^{(0)}(\omega\,;\sigma)=-\,\pi \left(\frac{2\pi\sqrt{\sigma}}{\sqrt{-\omega^2}}\right)^\frac{D-2}2\,Y_{\frac{D-2}2}\left(\sqrt{-\,\sigma\,\omega^2\,}\,\right)\,,\qquad (D \geqslant 2)
\label{Gsigma 0 Y}
\ee
where $D \geqslant 2$ and $Y$ is the Bessel function of the second kind.

\vspace{3mm}

Secondly, the generic generating functions $\texttt{G}^{(\ell)}$ can be obtained by combining the obtained equalities, respectively \eqref{Gsigma 0 K} and \eqref{Gsigma 0 Y}, with the last equality in \eqref{Gsigmaa}\,, to arrive at
\begin{equation}
\omega^2 > 0:~ \texttt{G}^{(\ell)}(\omega\,;\sigma)\,=\, 2 \, \pi^ {\,\ell} \left(\frac{\sqrt{\omega^2}}{2\pi\sqrt{\sigma}\,}\right)^ {\frac{2-D}{2} + \ell }\,K_{\ell-\frac{D-2}{2}}\left( \sqrt{\sigma\,\omega^2}\,\right) \,\cos\left[  \pi (\tfrac{2-D}{2} + \ell ) \right] \,,\label{Gsigma l K}
\end{equation}
for $D \leqslant 2 (\ell+1)$ and
\begin{equation}
\omega^2 < 0:~ \texttt{G}^{(\ell)}(\omega\,;\sigma)\,=\, -\,\pi^{1+\ell} \left(\frac{2\pi\sqrt{\sigma}}{\sqrt{-\omega^2}}\right)^{\frac{D-2}{2} - \ell  } \,Y_{\frac{D-2}{2} - \ell}\left(\sqrt{-\,\sigma\,\omega^2} \, \right)
 \,, \quad
\label{Gsigma l Y}
\end{equation}
for $D\geqslant 2(\ell+1)$\,.
Using \eqref{K-0} and \eqref{Y-0}\, we can obtain the limiting behaviour of the relations \eqref{Gsigma l K} and \eqref{Gsigma l Y} for small arguments of the (modified) Bessel functions, e.g. $\sigma\to 0^+$ or
\begin{eqnarray}
\omega^2 \to 0^+:~ \texttt{G}^{(\ell)}(\omega\,;\sigma) &\sim& \pi^ {\,\ell}~ \Gamma\le( \tfrac{2-D}{2}+\ell \ri)  \le( \tfrac{1}{\sqrt{\pi \, \sigma}} \ri) ^{2(\ell+1) - D} ~\cos\left[  \pi (\tfrac{2-D}{2} + \ell ) \right] \, , 
\label{Gsigma l K lim}
\\
\omega^2 \to 0^-:~ \texttt{G}^{(\ell)}(\omega\,;\sigma) &\sim& \pi^ {\,\ell}~\Gamma\le( \tfrac{D-2}{2}-\ell \ri) \le( \tfrac{2 \, \sqrt{\pi}}{\sqrt{-\, \omega^2}} \ri) ^{D - 2(\ell+1)}
\,.
 \label{Gsigma l Y lim}
\end{eqnarray}
The relation \eqref{Gsigma l K lim} is independent of $\omega^2$ and, hence, for non-zero $\sigma$\,, can be finite at the origin $\omega=0$.
However, due to the gamma function and cosine factors, for $D\geqslant 2$ the right-hand-side of \eqref{Gsigma l K lim} is actually infinite for $D$ even
and zero for $D$ odd. The relation \eqref{Gsigma l Y lim} is independent of $\sigma$ but it diverges at the origin $\omega=0$.
%
Note that, due to the two different limits on the dimension $D$ in the formulas \eqref{Gsigma l K} and \eqref{Gsigma l Y}, via an analytic continuation we can only try make a relation between them for $D=2(\ell+1)$. Using the identity \eqref{Analitic}, we get
\be
Y_0 (i\,z) \, =\, -\, \frac{2}{\pi} \, K_0 (z) + \frac{2}{\pi}\,
\le[ \log( i \, z ) - \log(z) \ri] ~ I_0(z)\,,
\ee
where $I$ is the modified Bessel function of the first kind and $z=\sqrt{\sigma\omega^2}$ in the present case. This shows explicitly that the relation between the timelike and spacelike cases is more subtle than a mere analytic continutation.

\subsection{Generating functions in Euclidean signature}\label{genfctsEucl}

To compare the generating function \eqref{Gsigma l Y} with the one which will be obtained in the Euclidean space, we extract the regular part of the latter in the relevant case (when $\omega$ is spacelike). Applying the definition of the Hankel functions of the first kind in terms of the modified Bessel functions of the second kind, the relation \eqref{Gsigma l Y} can be converted to
\be
\texttt{G}^{(\ell)}(\omega\,;\sigma)= i\,\pi^{\,1+\ell} \left(\frac{2\pi\sqrt{\sigma}}{\sqrt{-\omega^2}}\right)^{\frac{D}{2}-1 - \ell  }
 \le[ H^{(1)}_{\frac{D}{2}-1 - \ell}\left(\sqrt{-\sigma\omega^2}\right)  -\, J_{\frac{D}{2}-1 - \ell}\left(\sqrt{-\sigma\omega^2} \right)  \ri] \,.
\label{Gsigma l Y converted}
\ee
From the known properties of the Hankel and Bessel functions of the first kind, it is clear that the second term in the right-hand-side of the above expression is regular at the origin, for $D \geqslant 2 (\ell+1)$\,.

In order to avoid using distributions for the generating functions, the authors of \cite{Segal:2002gd,ST PRD} restricted their analysis to integrals over compact manifolds by introducing
the alternative generating functions \footnote{In the Euclidean space with Wick-rotated coordinates, a $(-i)$ prefactor has been omitted in the definition of the generating function, while one should take it into account for comparing to the Lorentzian solution \eqref{Gsigma l Y converted}\,.\label{footnoteref} }
\be
\textsc{G}^{(\ell)}_E(\omega;\sigma):= \int  d^D \bar\e \,\, \delta^{(\ell)} (\bar\e^2 +\sigma) \, e^{-i\, \bar\e \cdot \omega}=\left(\frac{\partial}{\partial\sigma}\right)^\ell \textsc{G}^{(0)}_E(\omega;\sigma)\,,\label{Gsigma}
\ee
where the integral is over the Euclidean space with
Wick-rotated coordinates,
 $$\bar\eta^\mu = (i\eta^0, \eta^1, \dots \eta^{D-1})\,,$$ so that
$$\bar\eta^2=\eta_{\mu\nu}\bar\eta^\mu\bar\eta^\nu=-\left[\,\left(\bar\eta^0\right)^2+\left(\bar\eta^1\right)^2+\dots+\left(\bar\eta^{D-1}\right)^2\,\right]\,.$$
The support of the distribution in \eqref{Gsigma} is the $(D-1)$-sphere in $D$-dimensional Euclidean space determined by $\bar\eta^2=-\sigma$, where $\sqrt{\sigma}$ is the radius.
Note that when $\omega$ is zero, the limit $\sigma \rightarrow 0$ corresponds to the case when the sphere shrinks to a point, thus $\textsc{G}^{(\ell)}_E(\omega;0)=0$. The definition \eqref{Gsigma} at $\ell=0$ implies
\be
\le( \p_\omega ^2 + \sigma\ri) \textsc{G}^{(0)}_E(\omega;\sigma) = \,0\,.\label{Helmequ}
\ee
Considering rotation symmetry $\textsc{G}^{(0)}_E(\omega;\sigma)$ only depends on $\omega$ through $r=\sqrt{- \omega^2}$, thus \eqref{Helmequ} becomes
\be
\le[ \frac{d^2 }{dr^2} + \frac{D-1}{r} \, \frac{d }{dr} + \sigma \ri] \textsc{G}^{(0)}_E (\omega;\sigma)=0\,.  \label{Int Bessel eq}
\ee
For $\sigma=1$, this equation reproduces 
 the Bessel differential equation \eqref{Bessel equation} with $\nu=\frac{D}{2} - 1$
in terms of the function $r^{\frac{D}{2}-1}\, \textsc{G}^{(0)}_E (r)$. Therefore, the solutions of (\ref{Int Bessel eq}) which are regular at the origin are only the Bessel functions of the first kind, of index $\frac{D}{2} - 1$. Combining this result with the last equality in \eqref{Gsigma} together with the recurrence relations \eqref{recurrence}, one finds
\be
\textsc{G}^{(\ell)}_E (\omega;\sigma) =  \pi^ {\, 1+\ell}\le(\frac{2\pi\sqrt{\sigma}}{r}\ri)^{{\frac{D}{2}} - 1 - \ell} \, J_{\frac{D}{2} - 1 - \ell}(\sqrt{\sigma}\,r)
\bigg|_{r=\sqrt{- \omega^2}} \,.   \label{Int Bessel func}
\ee
To write the latter, we used the normalisation fixed by the volume of the sphere for $\ell=0$. More generally, the value at the origin is:
\be
\textsc{G}^{(\ell)}_E (\omega;\sigma)\bigg|_{\omega=0}=\int \, d^D \bar\e \, \delta^{(\ell)} (\bar\e^2 +\sigma) =\frac{\pi^{\frac{D}2}
{\le(\sqrt{\sigma}\,\ri)}^{D - 2( \ell +1) }}{\Gamma\le(\tfrac{D}{2} - \ell\ri) } \,.
\ee
Notice that the integral over the $(D-1)$-sphere becomes zero in the limit $\sigma \rightarrow 0$ when $D > 2(\ell +1)$. In this limit, using \eqref{J limit}, the relation \eqref{Int Bessel func} will be proportional to $\sigma^{\frac{D}{2}-1-\ell}$ which yields zero, independently of the value of $\omega$\,.
However, as we discussed above, the Lorentzian solution was nonzero in the limit due to the presence of the Hankel function.
Indeed, the formula \eqref{Int Bessel func} is the regular part of \eqref{Gsigma l Y converted}, up to a factor $(-i)$ which comes from the definition of the generating functions (cf. footnote \ref{footnoteref}).

\section{Current exchange in the Euclidean signature}\label{eta_integrals}

By making use of the results of the appendix \ref{eta integrals} on the generating functions of the $\eta$-integrals in the Euclidean signature, one can check explicitly that the free actions of the Schuster-Toro type are not rank-diagonal.
We diagonalise these expressions by applying the Gegenbauer addition theorem.


In this work, we are interested to compute current exchanges obtained from (\ref{B S_int}) in $D$ dimensions. Therefore we have to compute integrals like
\be
I({\sigma}) := \int  d^D \e \, \delta' (\e^2 +\sigma) \, F_1(\e) \, F_2(\e) \,, \label{Int 1}
\ee
where the integration should be understood over the Euclidean signature, although we will omit the bar over $\eta$'s not to overload the notation. We want to express \eqref{Int 1} in terms of contractions of (traces of) the symmetric tensors appearing as coefficients in the two generating functions
\be
F_i (\e) = \sum_{s=0}^{\infty} \, \frac{1}{s!} ~ \e^{\m_{1}}   \ldots \e^{\m_{s}} ~\overset{(s)}{F_i} _{\,{\m_{1}\ldots\m_{s}}}
 \,, \quad\quad (i=1,2)\,.  \label{Int F_i}
\ee

As is stated in the preceding section \ref{genfctsEucl}, we encounter integrals like (\ref{Int 1}) which, using (\ref{Int F(etaa)}) and \eqref{Int Bessel func} with $\ell=1$, become
\bea
I({ \sigma}) &=&   \textsc{G}_E^{(1)}(i \,\partial_\eta;\sigma)  \,  F_1(\e) \, F_2(\e)  \,\bigg|_{\e=0}\,,  \\
&=& \pi^{D/2} \, \le(\tfrac{2 \sqrt{\sigma} }{ r}\ri)^{{\frac{D-4}{2}}} \, J_{\frac{D-4}{2}}\le( \sqrt{\sigma} \, r \ri) \,\, F_1(\e) \, F_2(\e) \,
\bigg|_{\e  =  0}
\,,  \label{Int 3}
\eea
where we introduced the pseudo-differential operator $r :=({\p_{\e}\cdot \p_{\e}})^{1/2}$\,.
We should take into account the Leibniz rule by rewriting \eqref{Int 3} as
\bea
I({ \sigma}) &=&   \textsc{G}_E^{(1)}\Big(i \,(\partial_{\eta_1}+\partial_{\eta_2}) \,\Big)  \,  F_1(\e_1) \, F_2(\e_2)  \,\bigg|_{\e_1=0=\e_2}\,,  \\
&=& \pi^{D/2} \, \le(\tfrac{2 \sqrt{\sigma} }{ \hat{r}}\ri)^{{\frac{D-4}{2}}} \, J_{\frac{D-4}{2}}\le( \sqrt{\sigma} \, \hat{r} \ri) \,\, F_1(\e_1) \, F_2(\e_2) \,\bigg|_{\e_1=0=\e_2}
\,,  \label{Inttt}
\eea
where the pseudo (bi)differential operator $\hat{r}$ must be understood as the expression
\be
\hat{r} = (r_1 ^2 + r_2 ^2 - 2\,r_{12})^\frac12 
 \,,
\ee
introducing the pseudo-differential operators $r_{12}:=\p_{\e_1}\cdot \p_{\e_2}$\,, $r_1 :=({\p_{\e_1}\cdot \p_{\e_1}})^{1/2}$ and $r_2 :=({\p_{\e_2}\cdot \p_{\e_2}})^{1/2}$\,.
Because of the presence of the pseudo-differential operators $r_1$ and $r_2$, the integral $I(\sigma)$ expressed in terms of the tensors \eqref{Int F_i} will not be rank-diagonal. More explicitly,
by expanding the Bessel function as the power series \eqref{Bessel}, we find (with $\sigma=1$)
\bea
I(1) & = & \sum_{n=0}^{\infty} \, g'_{n,D} \, \le( \p_\e \cdot \p_\e \ri)^n \, F_1(\e) \, F_2(\e) \,\bigg|_{\eta=0}   \label{Int 2} \\
& = &  \sum_{n=0}^{\infty}  \sum_{m=0}^{n} \sum_{k=0}^{n-m}   \tfrac{g'_{n,D}}{m!\,k!}\, \tfrac{  n! \, 2^m}{(n-m-k)!}
\underbrace{\le[ \p_\e ^m (\p_\e \cdot \p_\e)^{n-m-k} \, F_1(\e)  \ri] \cdot  \le[ \p_\e ^m (\p_\e \cdot \p_\e)^{k}   F_2(\e) \ri] \bigg|_{\eta=0}}_{=~\overset{(2n - m - 2k)} {F_1} {}^{~~ \m_1  \cdots \m_{n-m-k}} _{  {\m_1}  \cdots {\m_{n-m-k}}\, { \n_1} \cdots {\n_{m}} }
	\overset{( 2k + m)} {F_2} {}^{~~ \rho_1  \cdots \rho_{k}\, { \n_1 \cdots \n_{m} } } _{ {\rho_1  \cdots \rho_{k} } }} \,,  \nonumber
\eea
where $\p_{\e _\m}^m := \frac{\p}{\p \e^{\m_1}} \cdots  \frac{\p}{\p \e^{\m_m}} $\, and $g'_{n,D}:= \tfrac{(-1)^n}{2^{2 n} \, n!}\, \tfrac{\pi^{D/2}}{  \Gamma\le(n - 1 + D/2\ri)}$\,. This expression illustrates that, for example, the action (\ref{B S_int}) is not rank-diagonal.

One way to make the action rank-diagonal is by redefining the gauge fields.
In fact, by applying Gegenbauer's addition theorem \eqref{addition the.} for the Bessel functions, we can reobtain the redefined gauge fields of \cite{ST PRD}\,.
More precisely, one can first rewrite the relation (\ref{Inttt}) as
  \bea
I({ \sigma}) &=& \, \sigma^ {\frac{D-4}{2}} ~ \pi^{D/2}~2^{D-4}~ \Gamma(\tfrac{D-4}{2}) ~ \sum_{k=0} ^{\infty} \, \le(\tfrac{D-4}{2} + k\ri) ~ \mathcal{C}_k ^\frac{D-4}{2} \le( \tfrac{r_{12}}{r_1\, r_2} \ri)  ~
\le( \sigma \, r_1 \, r_2\ri)^{-\frac{D-4}{2}}\, \nonumber \\
& & ~ \times ~  J_{\frac{D-4}{2} + k}  \le(\sqrt{\sigma}\,r_1 \ri)   ~   J_{\frac{D-4}{2} + k}  \le( \sqrt{\sigma}\,r_2 \ri)  ~F_1
(\e_1)
~ F_2
(\e_2)
\,
 \Bigg| _{\e_1 \,=\, \e_{2} \,=\, 0}
 \,.  \label{Int 4}
 \eea
To get a more compact form, we can write the Bessel functions in \eqref{Int 4}, in terms of the hypergeometric functions \eqref{Bessel in Hyper}
\be
 J_{\frac{D-4}{2} + k}  \le( \sqrt{\sigma}\,r_i \ri) \, = \, \frac{\le( \sqrt{\sigma}\, r_i /2\,\ri)^{\frac{D-4}{2}+k}}{\Gamma\le(\frac{D-2}{2}+k\ri)}
  \,~  _0 \,F_1 \, \le( \, \tfrac{D-2}{2} + k \, ; \, -\, \sigma\, r_i ^2 / 4 \, \ri)  \,,  \quad\quad  i = 1, 2\,.
\ee
Then the relation \eqref{Int 4} becomes after some massage:
\be
I({ \sigma}) 
= \sigma^ {\frac{D}{2} - 2}~\sum_{k=0} ^{\infty} ~ A_{(\frac{D-4}{2},\, k)}
  ~ \le[ \,\widetilde{F_1}{_{\left(\frac{D-4}{2},\,k\right)}} (\e;\sigma)\, \ri] ~\, \overset{(k)}{\mathbb{P}_\frac{D-4}{2}}\, (\overset\leftarrow{\p}_\e, \overset\rightarrow{\p}_\e )~ \,
    \le[\, \widetilde{F_2}{_{\left(\frac{D-4}{2},\,k\right)}} (\e;\sigma) \,\ri]\, \Big| _ {\e \,=\, 0} \label{Int 4-4}
\ee
where the coefficient $A_{(\nu, k)}$ is given by
\be
A_{(\nu, k)}~ = ~ \frac{\pi^{D/2}}{ \,k! \, 2^{k} \,\Gamma(1+\n + k )\,}\,,
\label{A}
\ee
the operator $\mathbb{P}^{(k)}_\n$ (which we will call "spin-$k$ current-exchange operator") is defined as
\be
\overset{(k)}{\mathbb{P}_\n}\, (\overset\leftarrow{\p}_\e, \overset\rightarrow{\p}_\e )~ = ~ \frac{\, k! \,}{\, 2^k \, (\n)_k \,} ~
 ( \overset\leftarrow{\p}_\e \cdot \overset\leftarrow{\p}_\e )^{k/2}
~ \mathcal{C}_k ^\n \le(\tfrac{\overset\leftarrow{\p}_\e \cdot \overset\rightarrow{\p}_\e}
{\, (\overset\leftarrow{\p}_\e \cdot \overset\leftarrow{\p}_\e)^{1/2}
~ ( \overset\rightarrow{\p}_\e \cdot \overset\rightarrow{\p}_\e )^{1/2} \,} \ri) ~
 ( \overset\rightarrow{\p}_\e \cdot \overset\rightarrow{\p}_\e )^{k/2}\,,  \label{Current operator}
\ee
and $\widetilde{F_i}{_{(\n,k)}}(\e;\sigma)$ are redefined generating functions 
\bea
\widetilde{F_i}{_{\,(\n,k)}} (\e;\sigma) ~ &=& ~ {\sigma}^{\,k/2} ~  _0 \,F_1 \, \big( \,1+ \n + k\, ; \, -\, \sigma\, ( {\p}_\e \cdot {\p}_\e ) / 4 \, \big) ~ {F_i} \le(\e
\ri)\,.  \label{redefined fields}
\eea
 These redefined generating functions are written in terms of the hypergeometric function $_0 \,F_1$.
The symbol $(a)_n$ in the denominator of \eqref{Current operator} 
denotes the n-th Pochhammer symbol of $a$ \eqref{Pochhammer}\,.
By expanding the Gegenbauer polynomials \eqref{Gegenbauer}, the spin-$k$ current-exchange operator \eqref{Current operator} takes the form
\be
\overset{(k)}{\mathbb{P}_\n}\, (\overset\leftarrow{\p}_\e, \overset\rightarrow{\p}_\e )~ = ~
\sum_{n=0} ^{\lfloor k/2 \rfloor} \frac{1}{\,2^{2 n}\, (1-\n-k)_n } \,
\, \frac{k!}{ n! \, (k - 2 n)!} \,
(\overset\leftarrow{\p}_\e \cdot \overset\leftarrow{\p}_\e)^n \,
(\overset\leftarrow{\p}_\e \cdot \overset\rightarrow{\p}_\e)^{k-2n}\,
 ( \overset\rightarrow{\p}_\e \cdot \overset\rightarrow{\p}_\e )^n \,.  \label{Current operator 2}
\ee
Performing the evaluation at $\e = 0$ of \eqref{Int 4-4} and considering \eqref{redefined fields}
as generating functions, one finds (with $\sigma=1$)
\bea
& & 
I(1)
 ~=~ \sum_{s=0} ^{\infty} ~ 
\, A_{(\frac{D-4}{2}, s
)} ~ \times  \label{sigma = 1} \\
 & & \qquad\quad  \times ~ \sum_{n=0} ^{\lfloor s/2 \rfloor} \frac{1}{\,2^{2 n}\, (\,3 - \tfrac{D}{2}  - s \,)_n } \,
\, 
\frac{s!}{ n! \, (s - 2 n)!} ~~ {\overset{(s)}{\widetilde{F}_{\,1}}}{}^{[\,n\,]} \cdot {\overset{(s)}{\widetilde{F}_{\,2}}}{}^{[\,n\,]} \,, \nonumber
\eea
where $\widetilde{F}^{[\,n\,]}$ denotes the $n${th} trace of the tensor fields $\widetilde{F}$\, and the dot between the two tensors denotes contractions of the remaining indices.

\section{Useful functions and relations}\label{useful}

 In order to be self-contained, we collect here several useful definitions and formulas taken from \cite{Abramowitz}.

\vspace{1mm}
For any $n\in\mathbb N$ and $a\in\mathbb R$, the \textit{rising Pochhammer symbol} $(a)_n$ is defined as
\be
 (a)_n = \frac{\Gamma(a+n)}{\Gamma(a)} =
a \, (a+1) \cdots (a+n-1) \,.   \label{Pochhammer}
 \ee
A useful relation on the Euler gamma function is the reflection formula
\be
\Gamma(1-z) \, \Gamma(z) = \frac{\pi}{\, \sin (\pi\, z) \,} \, , \quad z\neq \mathbb{Z}\,.
\label{Gamma}
\ee

\vspace{1mm}
The \textit{double factorial} can be given in terms of the Euler gamma function as
\be
z!! \, =\, z (z-2) (z-4)\ldots (2\, \mbox{or} \,3) \,=\, \sqrt{\frac{2^{z+1}}{\pi}}  ~ \Gamma\le( 1 + \tfrac{z}{2} \ri)\,.  \label{Double}
\ee

\vspace{1mm}
The \textit{Bessel function of the first kind} $J_\n(z)$ is given as the power series
 \be
 J_\n(z) \, = \, \sum_{m=0}^{\infty} \, \frac{(-1)^m}{ \, m! \, \Gamma(m + \n +1)} ~ \le( \frac{z}{2} \ri)^{2m+\n} \,. \label{Bessel}
 \ee
The series \eqref{Bessel} converges for any finite $z$ and $\n$. It satisfies the differential equation
\be
\le( z^2 \, \p_z^2 \, + \, z \, \p_z \,+\, z^2 \, - \, \n^2 \,   \ri) \,  J_\n(z)\,=0 \,.  \label{Bessel equation}
\ee
The following useful relations, related to the Bessel function, were used in the text:
\be\label{recurrence}
\p_z \le( \, z^\n \, J_\n(z) \,\ri) \, = \, z^\n \, J_{\n-1} (z)  \,,
\ee
\be
\le( z \, \p_z^2 \, + \, (1+2\,\n)\, \p_z \,+\, z \,   \ri) \,  G_\n(z)\,=0 \,, \quad\mbox{where}\quad G_\n(z)  := z^{-\n} \, J_\n(z) \,,  \label{U}
\ee
\be
 \lim_{z \to 0} \le(~ z^{-\n} \, J_{\n+\rho}(z) ~\ri) \, = \, \left\{
                                                           \begin{array}{ll}
                                                             1 / (2^\n \, \Gamma(\n+1)\,)\,, & \hbox{\quad $\rho\,=\,0$\,;} \\
                                                             0\,, & \hbox{\quad$\rho> 0$\,.}
                                                           \end{array}
                                                         \right. \label{J limit}
\ee
An integral representation of the Bessel functions is given by
\be
J_{\nu}\left(z\right)=\frac{(\tfrac{1}{2}z)^{\nu}}{\pi^{%
\frac{1}{2}}\mathop{\Gamma\/}\left(\nu+\tfrac{1}{2}\right)}\int_{0}%
^{\pi}\, d\theta\,(\mathop{\sin\/}\theta)^{2\nu}\,  e^ {\mp i \,z\mathop{\cos\/}\theta} ~,\,\quad  \mathfrak{Re}(\nu)> -\tfrac{1}{2}\,.
 \label{Bessel int}
\ee
A useful integral containing the Bessel functions of the first kind is given in \cite{Vol. 2} as
\bea
& & \!\!\!\!\!\!\!\!\!\!\!\!\!\!\!\! \int_{a}^{\infty} \, dx ~~ x^{\,\n+1} ~ ( x^2 - a^2 )^{-\frac{1}{2}} ~ \exp{\big[{-\,b\, ( x^2 - a^2 )^{\frac{1}{2}} }\big]} ~ J_\n(x\,y) =
 -\, \le( \tfrac{\pi}{2} \ri)^{\frac{1}{2}} ~ ( a )^{\,\n+\frac{1}{2}}  ~  \times   \nonumber \\
& & \quad\qquad\quad\quad\quad\quad\quad \times   ~  {y^{\,\n}}~ {( y^2 + b^2 )^{-\frac{\n}{2} - \frac{1}{4} }  }
~~ Y_{\n+\frac{1}{2}} \big[ \,a \,( y^2 + b^2 )^{\frac{1}{2}} \big]  \,,    \label{J integral}
\eea
where $a, b>0$ and $\mathfrak{Re}(\nu)> - 1$\,.
The \textit{Gegenbauer's addition theorem} for the Bessel functions \cite{Vilenkin} reads
\be
\frac{ J_\n (\omega)}{\omega^\n} = 2^\n \, \Gamma(\n) \, \sum_{k=0} ^{\infty} \, (\n + k) \, \mathcal{C}_k ^\n (z)  ~
\frac{ J_{\n + k}  (x)}{x^\n}  ~ \frac{J_{\n + k}  (y)}{ y^\n} \,,  \label{addition the.}
\ee
where $\omega^2 = x ^2 + y ^2 - 2\, x\,y \,z $\,.

The \textit{Bessel function of the second kind} $Y_\n(z)$ is related to the one of the first kind by
\be
\mathop{Y_{\nu}\/}\nolimits\!\left(z\right)=\frac{\mathop{J_{\nu}\/}\nolimits%
\!\left(z\right)\mathop{\cos\/}\nolimits\!\left(\nu\pi\right)-\mathop{J_{-\nu}%
\/}\nolimits\!\left(z\right)}{\mathop{\sin\/}\nolimits\!\left(\nu\pi\right)}\,.
\ee
When $\nu$ is an integer the right-hand side is replaced by its limiting value. When $\nu$ is fixed\,,
\be
z\rightarrow 0 \,: \quad
\mathop{Y_{\nu}\/}\nolimits\!\left(z\right)\sim-(1/\pi)\mathop{\Gamma\/}
\nolimits\!\left(\nu\right)(\tfrac{1}{2}z)^{-\nu}\,, \quad \mathfrak{Re}(\nu)> 0 \,.\label{Y-0}
\ee
In addition, we have
\be
Y_\n(i\,z) \, =\, -\, \frac{2}{\pi \, i^{\,\n}} \, K_\n (z) + \frac{2\, i^{\,\n}}{\pi} \, \le[ \log(i\,z) - \log(z) \ri] \, I_\n (z) \,, \quad\quad \n \in \mathbb{Z}\,.
\label{Analitic}
\ee

\vspace{1mm}
The \textit{modified Bessel function of the first kind} $I_\n(z)$ is defined in terms of the Bessel function of the first kind as
\be
I_\nu(z)\,=\,i^{-\nu}J_\nu(iz)\,.  \label{J,I}
\ee
The \textit{modified Bessel function of the second kind} $K_\n(z)$ is defined in terms of the ones of the first kind as
\be
K_\n(z)\,=\,\frac{\pi}{2\,\sin(\nu\pi)}\,\Big(I_{-\nu}(z)-I_{\nu}(z)\Big)  \label{K sin}
\ee
When $\nu$ is fixed and $z\rightarrow 0$\,,
\be
\mathop{K_{\nu}\/}\nolimits\!\left(z\right)\sim\tfrac{1}{2}\mathop{\Gamma\/}%
\nolimits\!\left(\nu\right)(\tfrac{1}{2}z)^{-\nu}\,, \quad \mathfrak{Re}(\nu)> 0 \,. \label{K-0}
\ee
An integral representation of the latter is the identity
\be
K_\n(z) = \tfrac{1}{\sqrt{\pi}} ~ \le( \tfrac{2}{z} \ri)^\n ~ \Gamma(\n + \tfrac{1}{2})  ~
\int_{0}^{\infty}  dx  ~ \frac{~\cos ( x \,z) ~}{ (x^2 +1)^{\n + \frac{1}{2}}} \,,
\label{K}
\ee
where $ z > 0 $  and $\mathfrak{Re}(\nu)> -\tfrac{1}{2}$\,.
The following recurrence relations on the (modified) Bessel functions of the first and second kinds are useful:
\be
\p_z \le[ \, z^\n \, \mathcal{L}_\n(z) \,\ri] \, = \, z^\n \, \mathcal{L}_{\n-1} (z)
\ee
where $\mathcal{L}_\n$ denotes any of the functions $J_\n$, $Y_\n$, $I_\n$ and $e^{i\pi\n }K_\n$\,, and
\be
\frac{2\,\n}{z}\,\mathcal{L}_\n(z) = \mathcal{L}_{\n-1}(z) \pm  \mathcal{L}_{\n+1}(z)\,,
\qquad
\p_z \, \mathcal{L}_\n(z)  = \tfrac{1}{2}\, \le[ \mathcal{L}_{\n-1}(z) \mp \mathcal{L}_{\n+1}(z) \ri]
\ee
where $\mathcal{L}_\n$ denotes $J_\n$ and $Y_\n$ for the upper signs, and $I_\n$ and $e^{i\pi\n }K_\n$ for the lower ones.

\vspace{1mm}
The \textit{generalised hypergeometric function} $_p \,F_q
$ can be written as
\be
_p \,F_q  (\,a_1, \ldots, a_p\, ;\, b_1, \ldots, b_q \, ; \,z\,) \,= \, \sum_{n=0}^{\infty} \, \frac{\, (a_1)_n \ldots (a_p)_n \,}{(b_1)_n \ldots (b_q)_n} ~ \frac{z^n}{\,n!\,}
\ee
where $(a)_n$\,s are the Pochhammer symbol \eqref{Pochhammer}\,. The $z \rightarrow 0$ limit of this function is one.
The Bessel function can be expressed in terms of the hypergeometric function $_0 \,F_1$
\be
 J_{\n}  \le( z \ri) \, = \, \frac{(z /2)^{\n}}{\Gamma(\n+1)}
  \,~  _0 \,F_1 \, \le( \, \n  + 1\, ; \, -\, z^2 / 4 \, \ri)  \,,  \label{Bessel in Hyper}
\ee
with power series
\be
_0 \,F_1 \, \le( \, \n  + 1\, ; \, -\, z^2 / 4 \, \ri) \,=\,
\sum_{n=0}^{\infty} \, \frac{1}{ \, (\n  + 1)_n} \, \frac{( -\, z^2 / 4  )^n}{\, n!\,}    \,.  \label{Hyper series}
\ee

\vspace{1mm}
The \textit{Gegenbauer (or ultraspherical) polynomial} $\mathcal{C}_k ^\n  (z)$  with $\n>-\frac12$ and $\n\neq 0$ is a polynomial of degree $k\in\mathbb N$ in the variable $z$ defined as
\be
\mathcal{C}_k ^\n  (z) = \sum_{n=0} ^{\lfloor k/2 \rfloor}
 \frac{(-1)^n}{ n!} \,\, \frac{  \Gamma\le(\n + k - n \ri)}{ \Gamma\le(\n\ri) \, (k - 2 n )!\,}\, (2\,z) ^{k - 2 n} \,,
\label{Gegenbauer}
\ee
where $\lfloor k/2 \rfloor$ is the largest integer less than or equal to $k/2 $\,. Notice that $\mathcal{C}_0^\n (z)=1$\, and $\mathcal{C}^\n_k(-z)=(-1)^k \, \mathcal{C}^\n_k(z)\,$. The limit
\be
\mathcal{C}^0_k(z) \, =\, \lim_{\n \to 0} ~ \frac{1}{\, \n \,} ~ \mathcal{C}^\n_k(z)\,,
\ee
is related to Chebyshev polynomials of the first kind which can be used when we work in 4 dimensions.


\end{document}